\RequirePackage{booktabs}
\documentclass[pdflatex,sn-basic]{sn-jnl}
\usepackage{amsmath,amssymb,amsfonts}
\usepackage{amsthm}
\usepackage{adjustbox}
\usepackage{tabularx}

\usepackage{graphicx}
\graphicspath{Figures}

\usepackage{booktabs}
\usepackage{multirow}
\usepackage{float}
\usepackage{caption}
\usepackage{enumitem}

\usepackage{xcolor}
\definecolor{gray50}{gray}{.5}
\definecolor{gray40}{gray}{.6}
\definecolor{gray30}{gray}{.7}
\definecolor{gray20}{gray}{.8}
\definecolor{gray10}{gray}{.9}
\definecolor{gray05}{gray}{.95}
\definecolor{main}{HTML}{CFCFCF}
\definecolor{sub}{HTML}{CFCFCF}

\usepackage{listings}
\lstdefinestyle{mystyle}{
    language=Java,
    basicstyle=\ttfamily\small,
    keywordstyle=\color{blue},
    commentstyle=\color{gray},
    stringstyle=\color{red},
    breaklines=true,
    numbers=left,
    numberstyle=\tiny\color{gray},
    frame=single
}
\usepackage{algorithm}
\usepackage{algorithmicx}
\usepackage{algpseudocode}
\usepackage{fontawesome}
\usepackage[title]{appendix}

\usepackage{textcomp}

\newlength\Linewidth

\def\BibTeX{{\rm B\kern-.05em{\sc i\kern-.025em b}\kern-.08em
    T\kern-.1667em\lower.7ex\hbox{E}\kern-.125emX}}

\newcommand{\SATCountLetter}{three }
\newcommand{\RepoCount}{103 }

\newcommand{\ReviewerA}[1]{\textcolor{black}{#1}}
\newcommand{\ReviewerARev}[1]{\textcolor{black}{#1}}
\newcommand{\ReviewerB}[1]{\textcolor{black}{#1}}

\newenvironment{ReviewerAEnv}{\color{black}}{}
\newenvironment{ReviewerARevEnv}{\color{black}}{}
\newenvironment{ReviewerBEnv}{\color{black}}{}

\usepackage[many]{tcolorbox}
\tcbset{
    sharp corners,
    colback = white,
    before skip = 0.5cm,
    after skip = 0.7cm
}

\newtcolorbox{boxCNoTitle}{
    top=10pt,
    rounded corners,
    coltitle=black,
    colframe=gray,
    colback=sub,
    enhanced,
    center,
    boxrule=0.5pt
}

\newtcolorbox{boxC}[2][]{aibox,title=#2,#1}
\tcbset{
  aibox/.style={
    top=10pt,
    boxrule=0.5pt,
    rounded corners,
    coltitle=black,
    colframe=gray,
    colbacktitle=sub,
    colback=sub,
    enhanced,
    center,
    attach boxed title to top left={yshift=-0.1in,xshift=0.15in},
    boxed title style={boxrule=0.5pt,colframe=gray,},
  }
}

\newcounter{keyTakeAwaysCounter}

\newcounter{keyLimitationsCounter}

\newcounter{implicationsCounter}
\newenvironment{implications}[1][Implications]{
    \addtocounter{implicationsCounter}{1}
    \begin{boxC}{\faBook~\theimplicationsCounter. \textbf{#1}} 
}{
    \end{boxC}
}

\newcounter{keyRQAnswerCounter}

\begin{document}

\title{On the Correlation Between Architectural Smells and Static Analysis Warnings}

\author*[1]{\fnm{Matteo} \sur{Esposito}}\email{matteo.esposito@oulu.fi}
\author[1]{\fnm{Mikel} \sur{Robredo}}\email{mikel.robredomanero@oulu.fi}
\author[2]{\fnm{Francesca} \sur{Arcelli Fontana}}\email{francesca.arcelli@unimib.it}
\author[1]{\fnm{Valentina} \sur{Lenarduzzi}}\email{valentina.lenarduzzi@oulu.fi}

\affil*[1]{ \orgname{University of Oulu}, \orgaddress{\city{Oulu}, \country{Finland}}}

\affil[2]{ \orgname{University of Milano-Bicocca}, \orgaddress{\city{Milano}, \country{Italy}}}

\abstract{
\textit{Context.} Software quality assurance is essential during software development and maintenance. Static Analysis Tools (SAT) are widely used for assessing code quality. Architectural smells are becoming more daunting to address and evaluate among quality issues.

\noindent\textit{Purpose.} We aim to understand the relationships between Static Analysis Warnings (``warnings'') and Architectural Smells (``smell'') to guide developers/maintainers in focusing their effort on warnings more prone to co-occurring with smell. 

\noindent\textit{Methods.} We performed an empirical study on \RepoCount Java projects totaling 72 million LOC belonging to projects from a vast set of domains, and 785 warnings were detected by \SATCountLetter SAT, Checkstyle, Findbugs, PMD, SonarQube, and 4 architectural smells were detected by the ARCAN tool. We analyzed how warnings influence smell presence. Finally, we proposed an smell remediation effort prioritization based on warning severity and warning proneness to specific smells.

\noindent\textit{Results.} Our study reveals a moderate correlation between warnings and smells. Different combinations of SATs and warnings significantly affect smell occurrence, with certain warnings more likely to co-occur with specific smells. Conversely, 33.79\% of warnings are ``non-co-occurring'' with any of the smells in our dataset. \ReviewerB{This provides an early indicator for potential architectural concerns before resource-intensive architectural analysis is performed.}

\noindent\textit{Conclusion.} Practitioners can ignore about a third of warnings and focus on those most likely to be associated with smells. Prioritizing smell remediation based on warning severity or warning proneness to specific smells results in effective rankings like those based on smell severity. \ReviewerB{While not a substitute for specialized tools like ARCAN, warning-based prioritization provides a pragmatic bridge between low-level warnings and high-level architectural issues, particularly useful in contexts lacking full architectural visibility.}
}


 \maketitle

\section{Introduction}
\label{sec:Intro}
The risk of architectural erosion \cite{wan2023software,perry1992foundations} and disruption for critical services is exponentially increasing due to the pervasive nature of computing and the widespread adoption of digital services in all areas of daily life \citep{10356704}. To create a common definition for quality issues, the literature provides the concept of \textit{smells} \citep{10.1145/3345629.3345630}, referring to quality issues in code \citep{TaibiIST2017}. These smells are usually categorized as code smells, low-level coding issues \citep{Fowler1999,10.1145/3345629.3345630}, and, recently, architectural smells, higher-level abstract issues that concern the software architecture and the software components \citep{ArcelliFontana2017}.

A popular technique to detect quality issues is to leverage Static Analysis Tools (SAT)~\citep{vassallo2019developers, Avgeriou2020} which analyze the source code without executing it (thus static analyzers), and produce a report containing all the violations, i.e., Static Analysis Warnings (\ReviewerB{from now on ``warnings''}) that the code presents. More specifically, warnings refer to code violations with a lower-level granularity compared to smells. Therefore, warnings do not consider the software architecture of the system under development, commonly focusing on class or method levels \citep{esposito2024extensive}.

\ReviewerA{Despite the large body of work on smells detection techniques using architectural reconstruction, graph-based models, or clustering heuristics~\citep{mumtaz2021systematic}, these methods are often resource-intensive, demand explicit architectural views, or require specialized tooling that may not be readily integrated into developers’ daily workflows.}

\ReviewerA{In contrast, warnings are pervasive in software engineering practice. They are automatically produced, tightly coupled with development environments, and consulted regularly by practitioners. This makes warnings a pragmatic and low-barrier data source, yet their potential to act smells early indicators of architectural erosion remains largely underexplored.}

While multiple studies successfully exploited SATs to remediate common quality issues \citep{esposito2024extensive,Palomba2016,TaibiIST2017,LenarduzziJSS2020,ArcelliFontana2011d}, only a few studies focused on developing SATs to detect smells \citep{ArcelliFontana2017}. Moreover, some studies have investigated the interrelations between code anomalies~\citep{Macia2012,Macia2012b} or code smells~\citep{ArcelliFontana2019} and smells~\citep{Macia2012,Macia2012b}. However, to our knowledge, no previous study has analyzed the correlations between warning and smells.

Furthermore, to ensure the safety and stability of services, during development and maintenance, smell remediation should be prioritized~\citep{Nord2012,rachow2022architecture}. Hence, the portion of code affected by smells is the primary remediation target to prevent critical quality issues.

\ReviewerA{This work investigates whether warnings, widely available and frequently ignored, can serve as useful signals to identify or prioritize architectural remediation efforts. By doing so, we aim to bridge the gap between commonly available code-level signals and hard-to-detect architectural issues, providing practical insights for both researchers and practitioners.}

Therefore, the need to investigate how practitioners and researchers can exploit warnings to detect and remediate smells is evident. Hence, we designed and performed a large empirical study on the possible correlations between four smells and warning detected by \SATCountLetter tools largely adopted by developers \citep{esposito2024extensive}.

Our study assesses the correlation between warning and higher-level smells. We did not consider correlations between defects and warnings, smells, these have been extensively studied in the literature. Specifically, we investigate the correlation between warning and smells and how warning can prioritize smells.

Past studies have struggled to show an obvious relationship between warnings and smells \citep{pigazzini2021two}. Nonetheless, discovering a correlation between smells and warnings would enable the use of existing SAT smells as an effective tool for detecting issues in software architecture. \ReviewerA{Importantly, this approach requires no additional tooling or modeling effort, making it highly attractive in industrial contexts.} For practitioners, linking warning to architectural smells can reduce the need for refactoring, thereby saving both cost and effort~\citep{7544837,doi:10.1142/S021819402150008X}.

Therefore, we can summarize our main contributions as follows:
\begin{itemize}
	\item To our knowledge, we performed the first large-scale study, i.e., \RepoCount projects, on the correlation of smells and warnings. Other studies have considered correlations with more general architectural problems \citep{Macia2012b,Macia2012} or the correlation between smells and code smells \citep{ArcelliFontana2019}.
	\item We investigated the contribution of the most correlated warnings with smells.
	\item We proposed and evaluated smells remediation effort prioritization via warnings severity.
\end{itemize}

Our study focused on \RepoCount Java projects from the Qualitas Corpus dataset (QCD)~\citep{Tempero2010}, totaling 72 million lines of code analysis and  66,036 packages, 808,208 classes, and 88,460 interfaces examined across various contexts, including development platforms, multimedia, gaming, middleware, compilers, and testing frameworks. We considered a large set of warnings (1142) and four smells based on dependency issues that can critically influence the quality of a software project and its progressive architecture erosion \citep{Arcan-ICSME2016,li2021erosion,MUMTAZ2021110885}.

Contrary to past studies, our findings reveal a moderate, yet statistically significant, correlation between warnings and smells. Moreover, different combinations of SATs and warnings significantly influence the occurrence of smells. Therefore, specific warnings are more prone to co-occur with specific smells. \ReviewerA{Conversely, we also identified that warnings non-co-occurring with any smells represent 33.79\% of our dataset.} Our study shows that practitioners can safely drop a third of the warnings in our context and focus on the most smell-prone areas to address smells without prior smell severity or discovery knowledge effectively. Our study highlighted that prioritizing smell remediation efforts based on warning severity, or warning proneness to specific smells with a limited inspection window, results in rankings similar to the optimal one, i.e., ordered by smell severity.
Therefore, we suggest that warnings can moderately influence the presence of smells. Prioritizing warnings based on occurrence probability effectively guides remediation. Future research should improve the detection of smell-prone warnings and explore deeper connections between warnings and smells for better understanding and prioritization of remediation efforts.

\noindent\textbf{Paper Structure}. Section~\ref{sec:Background} describes the background to which our paper is based. In Section~\ref{sec:CS}, we present the study design. In Section~\ref{sec:Results}, we show the results obtained and discuss them in Section~\ref{sec:Discussion}. Section~\ref{sec:Threats} focuses on threats to the validity of our study. Section~\ref{sec:RW} discusses related work and in Section~\ref{sec:Conclusion}, we draw conclusions and outline future research directions. 

\noindent\ReviewerB{\textbf{Definitions:} In the remainder of this paper, we refer to static analysis warnings as ‘warnings’ and architectural smells as ‘smells’.}
\section{Background}\label{sec:Background}
This section introduces the static analysis tools used, the types of architectural smells considered, and the effort-aware metrics applied to evaluate prioritization. 

\subsection{Static Analysis Tools and Warnings}
This section introduces the tools we employed in our study and the motivation for their selection. 
SAT analyzes software without running it by inspecting the source code, thus discovering potential quality issues in the code base~\citep{Ernst2015}. Recently, the readily availability of SATs and the subsequent inclusion in Continuous Improvement (CI) practices~\citep{zampetti2017open} are boosting their popularity. Among the available SATs, we selected the most widely used according to state of the art in detecting warning~\citep{vassallo2019developers,esposito2024extensive} such as smells, Sonarqube, Checkstyle, Findbugs, and PMD. According to architectural issues, our study focuses on architectural smells. Therefore, we also included ARCAN based on our previous experience~\citep{ArcelliFontana2017}. In Table~\ref{tab:static_analysis_summary}, we summarize the characteristics of each of the following SATs. 
\begin{table*}[htbp]
\centering
\small
\resizebox{\linewidth}{!}{%
\begin{tabular}{p{2.5cm}|p{2.5cm}|p{2.5cm}|p{2.5cm}|p{2.5cm}}
\hline
\textbf{SAT} & \textbf{Checkstyle} & \textbf{Findbugs} & \textbf{PMD} & \textbf{SonarQube} \\ \hline
\textbf{Warnings Types} & 14 types & 9 types & 8 types & 3 types \\ \hline
\textbf{Total Warnings} & 173 & 424 & 305 & 413 \\ \hline
\textbf{Types} & Annotations, Block Checks, Class Design, Coding, Headers, Imports, Javac Comments, Metrics, Miscellaneous, Modifiers, Naming Conventions, Regexp, Size Violations, Whitespace & Bad practice, Correctness, Experimental, Internationalization, Malicious code vulnerability, Multithreaded correctness, Performance, Security, Dodgy code & Best Practices, Code Style, Design, Documentation, Error Prone, Multithreading, Performance, Security & Bugs, Code Smells, Vulnerabilities \\ \hline
\textbf{Severity Levels} & Error, Ignore, Info, Warning & 1-20 ranking: Scariest, Scary, Troubling, Concern & Change required, Change suggested, Change optional, Change highly recommended, Change highly optional & Blocker, Critical, Major, Minor, Info \\ \hline
\end{tabular}%
}
\caption{Summary of Static Analysis Tools Warning (warnings)}
\label{tab:static_analysis_summary}
\end{table*}

More specifically, regarding SATs, \textbf{Checkstyle}\footnote{\url{https://checkstyle.org}}
\label{Checkstyle} is designed to asses Java code quality.  The tool uses ``rules''  according to a set of \textit{checks} for analyzing the code base. Checkstyle has two predefined rule configurations: Google Java Style and Sun Java Style. 
Static violations of the checks are grouped under two severity levels: error and warning. 

\textbf{FindBugs}\footnote{\url{http://findbugs.sourceforge.net}} \label{FindBugs} examine Java bytecode.  The tool detects \textit{bug patterns} which are caused by, but not limited to, difficult language features and misunderstood API features.
Bug patterns are ranked on an ordinal scale from one to 20. Rank 1-4 is the \textit{scariest} group, rank 5-9 is the \textit{scary} group, rank 10-14 is the \textit{troubling} group, and rank 15-20 is the \textit{concern} group. 

\textbf{PMD}\footnote{\url{https://pmd.github.io/}} \label{PMD} can inspect eight programming languages, including  Java. PMD uses a set of \textit{rules} to assess code quality. There are a total of 33 different rule-set configurations for Java projects.   Rules are classified into 8  categories. The tool measures rule violations on an ordinal scale of one to five. One is associated with the most severe violations, and 5 is the least severe. 

\textbf{SonarQube}\footnote{\label{Sonar}\url{http://www.sonarsource.org/}} \label{SonarQube} is one of the most common SATs for detecting code quality issues.  SonarQube measures several aspects of the source code, such smells several lines of code and code complexity, and verifies code compliance with a specific set of \textit{coding rules} defined for most common programming languages. Violations of coding rules are reported smells \textit{issues} in  SonarQube. The estimated time to remove these issues, i.e., remediation effort, is used to compute the remediation cost. SonarQube analysis encompasses quality aspects such smells reliability, maintainability, and security.  Reliability rules, also named \textit{Bugs}, create quality issues that ``represent something wrong in the code'' and that will soon be reflected in a bug.  \textit{Code smells} are considered  ``maintainability-related issues'' in the code that decrease code readability and modifiability. It is important to note that the term ``code smells'' adopted in SonarQube does not refer to the commonly known code smells term defined by Fowler et al.~\citep{Fowler1999}, but to a different set of rules.

\begin{table}[htb]
	\centering
	\begin{tabular}{l | lll}
		\hline
		\multirow{2}{*}{\textbf{Tool}} & \multicolumn{3}{c}{\textbf{Detection Capability}} \\ \cline{2-4} 
		           & \textbf{\# rule} & \textbf{\# type} & \textbf{\# severity} \\ \hline
		Checkstyle & 173              & 14               & 4                    \\
		Findbugs   & 424              & 9                & 4                    \\
		PMD        & 305              & 8                & 5                    \\
		SonarQube  & 413              & 3                & 5                    \\ \hline
	\end{tabular}%
	\caption{Static Analysis Warnings}
	\label{tab:Warnings}
\end{table}

\subsection{Architectural Smells}
On the other hand, we employed \textbf{ARCAN }\footnote{\url{https://www.arcan.tech}} that detects different smells based on dependency issues by computing different metrics proposed by Martin~\citep{Martin1995}, such smells those related to instability issues. We refer to software instability smells the inability to make changes without impacting the entire system or a large part. 

The tool relies on graph database technology to perform graph queries, which enables higher scalability during the detection process and management of many dependencies.
The detailed detection techniques for smells have been described in previous works~\citep{ArcelliFontana2017}.  Moreover, the tool allows the computation of an architectural debt \citep{Roveda2018} index based on the number of smells detected in a project and the criticality of the smells, evaluated according to each type of smell.
The tool's results were validated on ten open source systems and two industry projects with a high precision value of 100\% in the results and 63\% recall \citep{ArcelliFontana2017}.
Moreover, the results of ARCAN were validated using practitioners' feedback on four industry projects~\citep{Martini2018}.
A research version of the ARCAN tool has been used for this study, while the latest Arcan commercial version is available at ~\footnote{\href{https://www.arcan.tech/}{https://www.arcan.tech/}}.

In our study, we have considered the following smells:
\textbf{Unstable Dependency (UD)}: An \textit{Unstable Dependency} occurs when a stable component depends on a less stable component. ~\citep{Martin2007}. 
This can lead to a ripple effect when the unstable component changes. ~\citep{ArcelliFontana2017}. Detected in packages.
\textbf{Example:} A stable \texttt{PaymentProcessor} component depending on an unstable \texttt{BetaFeature}.

\begin{lstlisting}
class BetaFeature {
    public void experimentalMethod() { /* Unstable feature */ }
}

class PaymentProcessor {
    private BetaFeature feature; // Depends on unstable module

    public PaymentProcessor(BetaFeature feature) {
        this.feature = feature;
    }
}
\end{lstlisting}

\textbf{Hub-Like Dependency (HD)}: A \textit{Hub-Like Dependency} occurs when an abstraction has many incoming and outgoing dependencies, making it a critical component that is difficult to modify.~\citep{Suryanarayana2015}. Detected in classes and packages.

\textbf{Example:} A \texttt{DatabaseHelper} class used across multiple services.

\begin{lstlisting}
class DatabaseHelper {
    public void connect() { /* Connect to DB */ }
    public void executeQuery(String query) { /* Run Query */ }
}

class UserService {
    private DatabaseHelper dbHelper = new DatabaseHelper();
}

class ProductService {
    private DatabaseHelper dbHelper = new DatabaseHelper();
}
\end{lstlisting}

\textbf{Cyclic Dependency (CD)}: A \textit{Cyclic Dependency} occurs when multiple components depend on each other in a cycle, making independent modifications and releases difficult.
The subsystems involved in a dependency cycle can hardly be released, maintained, or reused in isolation. 
Detected in classes and packages. The \textit{Cyclic Dependency} smells is detected according to different shapes~\citep{DBLP:conf/aswec/Al-MutawaDMM14} smells described in~\citep{ArcelliFontana2017}.
\textbf{Example:} A dependency cycle between \texttt{A}, \texttt{B}, and \texttt{C}.

\begin{lstlisting}
class A {
    private B b;
}

class B {
    private C c;
}

class C {
    private A a; // Cycle formed: A -> B -> C -> A
}
\end{lstlisting}

We focused on the above smells based on dependency issues since dependencies are very important in software architecture. Components with many dependencies can be considered more critical and expensive to maintain. Moreover, we defined smells ``non-co-occurring'' the instances without smells, i.e.,  a software entity affected by one or more warnings but without any of the ASs considered in our study.

\subsection{Effort Aware Metrics}
\label{subsec:eam}
\citet{ccarka2022effort} investigated effort-aware metrics (EAM) to improve defect prioritization. In our study, we take inspiration from the definition of PofB and PopT to create three different prioritization rankers for smells. We then evaluate our ranker in a way that is similar to the interpretation of the PofB results. Therefore, we briefly discuss EAMs; according to \citet{ccarka2022effort}, two different EAMs exist: normalized by size or not by size. The most known not-normalized EAM is called \textit{PofB} \citep{DBLP:journals/jsjkx/ChenLG0Z18, DBLP:journals/tse/WangLNT20, DBLP:journals/tse/XiaLPNW16,9064604} PofB, or Proportion of Defects Identified by the top x\% of the codebase, is a measure derived from a prediction model's rankings. It signifies the percentage of defective entities discovered when analyzing the initial x\% of the code. Higher PofB indicates more effective rankings, thereby enhancing testing support. For instance, a PofB10 of 30\% implies that 30\% of defective entities were detected by analyzing 10\% of the codebase according to the method's ranking.

Comparing a prediction model's ranking with a perfect one sheds light on its performance. Mende et al.~\citep{DBLP:conf/promise/MendeK09}, drawing inspiration from Arisholm et al.~\citep{DBLP:conf/issre/ArisholmBF07}, introduced "Popt" to gauge this accuracy. It assesses how the prediction model deviates from perfection while outperforming random rankings. Popt quantifies the area ${\Delta opt}$ between the ideal and predicted models. In the optimal scenario, instances are ordered by decreasing fault density, while in the predicted model, they're ordered by decreasing predicted defectiveness. The equation for Popt, ${1 - \Delta opt}$, indicates a larger value denotes a smaller gap between the ideal and predicted models~\citep{DBLP:journals/tse/YuWHH19}.

Popt and PofB are distinct metrics, each capturing different facets of model accuracy. They employ different ranking methods: Popt ranks entities based on bug density (bug probability divided by entity size), while PofB ranks them according to bug probability. Consequently, classifiers ranked by Popt and PofB may differ. Popt, grounded in bug density, offers a more realistic perspective than PofB, which relies solely on probability. However, interpreting Popt is more challenging; a classifier with double the Popt value doesn't necessarily yield double the benefits to the user.

\section{Empirical Study Design}
\label{sec:CS}
This section presents the empirical study's goal, research questions, metrics, and hypotheses. Moreover, we describe the study context, the data collection, and the data analysis procedures.
We designed and conducted our empirical study according to the guidelines defined by Wohlin et al.~\citep{DBLP:books/daglib/0029933}. 

\subsection{Goal, Research Questions, Metrics, and Hypothesis}
We formalized the \textbf{goal} of this study according to the Goal Question Metric (GQM) approach~\citep{Basili1994}  as follows:

The \textit{goal} of our empirical study is to investigate the relationship between architectural smells and static analysis warnings in the context of software quality assurance. Our \textit{perspective} is of practitioners and researchers seeking to assess the relationship between high-level architectural issues and low-level code quality problems.

Based on our goal, we formulate the following Research Questions (RQ$_s$)

\begin{boxC}{RQ$_1$.}  
	Are warnings and smells correlated?
\end{boxC}

Investigating the correlation between smells and warnings enables us to assess the relationship between high-level architectural issues and low-level code quality problems. We aim to determine whether smells are correlated with warnings and the strength of this correlation. To address RQ$_1$, we collected warning and smell presence per software package. \ReviewerB{To assess statistical significance, we conjectured a family of hypotheses, one for each warning-smells pair \textbf{hypothesis} ($H_1$)  as follows:}

\begin{ReviewerBEnv}
	    
	\begin{itemize}[labelindent=1.5em,leftmargin=5em]
		\item[$H_{11(w, s)}$] \textit{There is a correlation between the number of instances of warning $w$ and the number of instances of architectural smell $s$, averaged at the package level.}
	\end{itemize}
	
	\noindent Hence, we defined the \textbf{null hypothesis} ($H_0$)  as follows:
	\begin{itemize}[labelindent=1.5em,leftmargin=5em]
		\item[$H_{01(w, s)}$] \textit{There is no correlation between the number of instances of warning $w$ and the number of instances of architectural smell $s$, averaged at the package level.}
	\end{itemize} 
\end{ReviewerBEnv}

Moreover, different static analyzer warnings can lead to specific architectural smells.  Therefore, assessing whether a specific warning's presence is a possible indicator of the presence of specific smells is essential. Therefore, we ask:

\begin{boxC}{RQ$_2$.} 
	\ReviewerB{Do specific types of warnings co-occur with specific types of smells?}
\end{boxC}

In the long run, architectural smell can silently deteriorate complex software architecture. Hence, we investigate the characteristics of warnings affecting software projects and leading to smells to identify common traits among projects prone to smells and assess further factors contributing to the emergence and persistence of smells. To address RQ$_2$, we analyzed the co-occurrence of warnings and smells. Let $\sigma$ be a warning and $\alpha$ be a smell, and $\eta$ be a software entity. We state that $\sigma$ co-occur with $\alpha$, if given $\eta$ \textit{such that} $\sigma$ and $\alpha$ affect contemporary $\eta$. To assess statistically significance, we conjectured two \textbf{hypotheses} ($H_2$ and $H_3$ )  as follows:

\begin{ReviewerBEnv}
	    
	\begin{itemize}[labelindent=1.5em,leftmargin=5em]
		\item[$H_{12{(w,s)}}$] \textit{There is a difference in the co-occurrence of warning $w$ with smell $s$ compared to its co-occurrence with other smells.}
		\item[$H_{13{(w,s)}}$] \textit{There is a difference in the co-occurrence of warning $w$ with smell $s$ compared to its co-occurrence with other smells by SAT.}
	\end{itemize}
	
	\noindent Hence, we defined the \textbf{null hypotheses} ($H_0$)  as follows:
	\begin{itemize}[labelindent=1.5em,leftmargin=5em]
		\item[$H_{02{(w,s)}}$] \textit{There is no difference in the co-occurrence of warning $w$ with smell $s$ compared to its co-occurrence with other smells.}
		\item[$H_{03{(w,s)}}$] \textit{There is no difference in the co-occurrence of warning $w$ with smell $s$ compared to its co-occurrence with other smells by SAT.}
	\end{itemize} 
	
\end{ReviewerBEnv}

We computed the co-occurrence of each warning and smell. For each project package, we counted the number of non-co-occurring warnings (\textbf{NCO}) and those affected by CDs, UDs, and HLs (and combinations of them). We tested $H_2$ and $H_3$ with the Wilcoxon Paired Signed Rank Test (WPT) (Section \ref{sec:DataAnalysis}).

Finally,  assessing the correlation between warnings and smells and understanding how this relationship influences the presence of smells in software projects merely scratches the surface. Indeed, it falls short, considering practitioners' limited time for remediation efforts. Therefore, prioritizing these efforts becomes paramount. Thus, we ask:

\begin{boxC}{RQ$_3$}
	Are warnings effective for smell remediation effort prioritization?
\end{boxC}

Effort prioritization is essential to avoid wasting resources on noncritical tasks \citep{jorgensen2006systematic}. No prior study has investigated smell prioritization via warning severity.
According to the ``no free lunch theorem'', currently, there is no default severity for code quality checks universally accepted by our community \citep{esposito2023can}. Therefore, in our study, we investigated exploiting warning and warning attributes detected from SAT, such as smells SonarQube, i.e., rule severity and cost estimation,  and the prioritization effort for smell severity according to \citet{ArcelliFontana2017}.  We referred to recent empirical studies to assess the accepted severity of smells \citep{fontana2020architectural,pigazzini2021two,sas2023architectural}. More specifically, \citet{sas2023architectural} proposes, on average, an ordinal scale of one to ten, where one is the least severe smell and ten is the highest. \ReviewerA{Therefore, we adopt the average smell severity values, CD (5), UD (7), and HL (9), smells proposed by \citet{sas2023architectural} who empirically derived these ratings based on practitioner feedback and analysis of the impact of smells on maintainability and evolution.} Moreover, regarding warning severity, we compiled a comprehensive dataset in our previous study \citep{10.1145/3345629.3345630} focused on SonarQube. Without the risk of a generalizability issue \citep{esposito2023can}, we focused this RQ on SonarQube-specific warnings. To address RQ$_3$,  we quantify the \textbf{probability of a warning inducing smells} ($\mathcal{P}$) smells the relative frequency at which the particular warning contributed to the occurrence of smells. To assess statistical significance, we conjectured one \textbf{hypothesis} ($H_4$)  as follows:

\begin{itemize}
	\item[$H_{14}$] \textit{There is a  difference in the prioritized effort between the three rankers.}
\end{itemize}

\noindent Hence, we defined the \textbf{null hypothesis} ($H_0$)  as follows:
\begin{itemize}
	\item[$H_{04}$] \textit{There is no  difference in the prioritized effort between the three rankers}
\end{itemize} 
We computed $\mathcal{P}$ for each SAT, warning, and smell. We ranked the warnings based on their severity \citep{10.1145/3345629.3345630}, $\mathcal{P}$, and smells severity \citep{sas2023architectural}. We tested $H_4$ with WPT (see Section \ref{sec:DataAnalysis}).

\subsection{Study Context}
\label{context}
We selected projects from the Qualitas Corpus (QCD) collection of software systems~\citep{Terra2013}, using the compiled version of the QCD~\citep{Tempero2010}. Table \ref{tab:QCmetrics} details metrics on the QCD.

\begin{table*}
\centering
\small
\begin{tabular}{p{3cm}|p{8cm}}
\hline
\textbf{Metric} & \textbf{Value} \\
\hline
Number of Projects & 112 \\
Lines of Code (LOC) & $>$ 18 million \\
Packages & 16,509 \\
Classes Analyzed & 202,052 \\
Interfaces Analyzed & 22,115 \\
Methods Analyzed & 464,893 \\
Contexts Covered & IDEs, SDKs, databases, 3D/graphics/media, diagram/visualization libraries and tools, games, middlewares, parsers/generators/make tools, programming language compilers, testing libraries, and tools. \\
Available Metrics & LOC, NOP, NOCL, NOI, NOM, NOA, NORM, PAR, NSM, NSA. CK Metrics: WMC, DIT, NOC, LCOM HS. - MLOC, SIX, VG, NBD, RMD. - CA, CE, I, A. \\
\hline
\end{tabular}
\caption{Summary of Qulitas Corpus Metrics}
\label{tab:QCmetrics}
\end{table*}

\subsection{Study setup and data collection}

\begin{figure*}
	\centering
	\includegraphics[width=\linewidth]{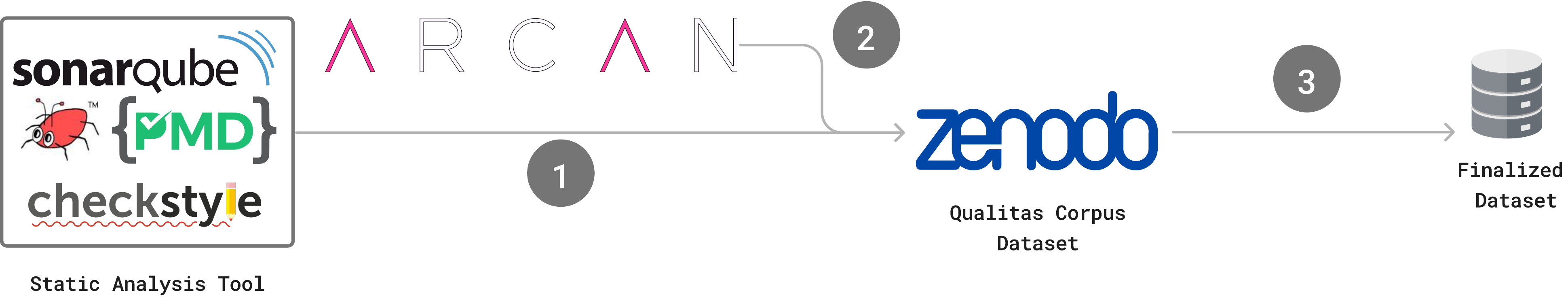}
	\caption{Data Collection Workflow}
	\label{fig:workflow}
\end{figure*}

This section presents our data collection methodology. Figure \ref{fig:workflow} presents the study workflow. Our data collection comprised three steps: 
\begin{enumerate}
	\item We analyzed the QCD with \SATCountLetter SATs and averaged the results for project packages;
	\item We separately analyzed QCD with the Arcan tool;
	\item To get our study results, we analyzed the finalized dataset, comprising all warning detected by the three SATs and the smells detected by the Arcan tool grouped by software package (see Section \ref{sec:DataAnalysis}).
\end{enumerate}

We analyzed the entire data set (RepoCount Java projects) from the Qualitas Corpus~\citep{Terra2013}. %However, we obtained results for only \RepoCount projects because the computational time for the remaining ones would have been excessively long due to project stratucture and size.

\subsection{Data Analysis}
\label{sec:DataAnalysis}

We considered the packages affected by at least a smell and one static analysis warning at the class or package levels.
One or more static analysis warnings could affect the same set of packages for each smell. This is because an smells may involve more than one class, while static analysis warnings involve only one class.
In the case of projects not infected by static analysis warnings or smells, we did not consider them for the analysis.

To answer \textbf{RQ$_1$}, we collected the QCD projects and ran the SAT and ARCAN. We merged the outputs of the tool. 
In our context, the dependent variable is the smell, and the independent variables are the warning.
To compute the correlation coefficients, we need to assess the distributions of our variable. Therefore we conjectured one \textbf{hypothesis} ($H_\mathcal{N}$)  as follows:
\begin{ReviewerBEnv}
	    
	\begin{itemize}[labelindent=1.5em,leftmargin=5em]
		\item[$H_{1\mathcal{N}}$] \textit{The collected warning and smells metrics do not follow a normal distribution.}
	\end{itemize}
	
	\noindent Hence, we defined the \textbf{null hypothesis} ($H_0$)  as follows:
	\begin{itemize}[labelindent=1.5em,leftmargin=5em]
		\item[$H_{0\mathcal{N}}$] \textit{The collected warning and smells metrics follow a normal distribution.}
	\end{itemize} 
\end{ReviewerBEnv}

\begin{ReviewerBEnv}
	We tested for normality ($H_\mathcal{N}$) using the Anderson–Darling (AD) test \citep{doi:10.1080/01621459.1974.10480196}, considered one of the most powerful methods for detecting deviations from normality \citep{stephens2017tests}. Since normality could not be assumed, we used Spearman’s $\rho$ \citep{spearman_rho} instead of Pearson’s correlation coefficient \citep{pearson1895notes}.
	
\end{ReviewerBEnv}
Table \ref{tab:spearmaninterpretation} presents $\rho$ values interpretation according to \citet{dancey2007statistics}.
\begin{table*}
\centering
\small
\resizebox{\linewidth}{!}{%
\begin{tabular}{l|r|rrr|rrr|rrr|r}
\hline
\textbf{Interpretation       }        & \multicolumn{1}{r}{\textbf{Perfect}} & \multicolumn{3}{|c|}{\textbf{Strong}} & \multicolumn{3}{c|}{\textbf{Moderate}} & \multicolumn{3}{c|}{\textbf{Weak}} & \multicolumn{1}{r}{\textbf{Zero}} \\ \hline
\multirow{2}{*}{\textbf{Correlation}} & + 1                         & + 0.9   & +0.8    & + 0.7  & +0.6     & + 0.5   & + 0.4   & + 0.3  & + 0.2  & + 0.1  & 0                        \\
                             & - 1                         & - 0.9   & - 0.8   & - 0.7  & - 0.6    & - 0.5   & - 0.4   & - 0.3  & - 0.2  & - 0.1  & 0                        \\ \hline
\end{tabular}%
}
\caption{Spearman's $\rho$ interpretation (RQ$_1$)}
\label{tab:spearmaninterpretation}
\end{table*}
\begin{ReviewerBEnv}
	Therefore, we tested $H_{01(w, s)}$ with Spearman's $\rho$ for each warning-smells pair, i.e., warning $w$ and smell $s$, resulting in 663 distinct tests. Each test measured the strength of correlation between the number of instances of $w$ and $s$ at the package level. 
\end{ReviewerBEnv}

To answer \textbf{RQ$_2$}, we computed Spearman's $\rho$ for each warning-smells combination.  We used the interquartile ratio (IQR) method  \citep{hojo1931distribution} to select the top correlated warning for each SAT and smells, i.e., we computed the quartiles. We selected only the warnings with their Spearman's $\rho$ value higher than or equal to the third quartile (Q3), thus selecting the top 25\%. We computed the $\mathcal{P}$  as follows:
Let \( \Sigma_{\text{SAT}_i} \) represent the set of Static Analysis Warnings specific to a particular Static Analysis Tool \( \text{SAT}_i \). Then, \( \sigma \in \Sigma_{\text{SAT}_i} \) signifies that \( \sigma \) belongs to the set of warnings specific to \( \text{SAT}_i \); thus \( \sigma_{j} \) is a specific warning and \( \alpha \in smells \) represents a specific smells.. Therefore, we define the relative frequency \( \mathcal{P}(\sigma_{j}, \alpha) \) smells:

\[ \mathcal{P}(\sigma_{j}, \alpha) =  \frac{| \sigma_{j} \text{ in } \alpha|}{| \sigma \text{ in } \alpha |}\]

Where:
\begin{itemize}
	\item The numerator represents the number of times the specific warning, \( \sigma_{j} \in \Sigma_{\text{SAT}_i} \), appears within the given smells, \( \alpha \).
	\item The denominator represents the total count of all warnings belonging to a specific SAT, \( \sigma \), in the given smells, \( \alpha \).
\end{itemize}

\ReviewerB{Informally, $P(\sigma_j, \alpha)$ expresses how likely a specific warning $\sigma_j$ is to appear within the context of packages affected by a given smell type $\alpha$. This helps identify which warnings are most frequently associated with particular architectural issues.
}

More specifically, we computed $\mathcal{P}$ for each possible combination of smells. Therefore, our data computed a $\mathcal{P}$-value for the 3 smells, the non-co-occurring instances, and for combinations of smells.
We tested $H_{2}$, and $H_{3}$ via the WPT \citep{Conover}, which is a non-parametric statistical test that compares two related samples or paired data. WPT uses the absolute difference between the two observations to classify and then compare the sum of the positive and negative differences. The test statistic is the lowest of both. We selected WPT to test $H_{2}$ and $H_{3}$ because the data was not normally distributed; hence, we used it instead of the paired t-test, which assumes a normal data distribution.

Moreover, due to the high number of statistical tests performed, we needed to apply p-value correction. \ReviewerB{In place of the usual level of significance ($\alpha=0.05$), we need to correct p-values since it is essential in multiple hypothesis testing to control the increased risk of Type I errors.} When multiple tests are conducted simultaneously, the probability of incorrectly rejecting at least one true null hypothesis increases. A classical p-value correction technique is the Bonferroni correction. However, Bonferroni is known to offer a conservative correction~\citep{hochberg1988sharper}. Therefore, we decided to use the Benjamini \& Hochberg (BH) p-value correction~\citep{benjamini1995controlling}, which uses the false discovery rate (FDR) when performing multiple hypothesis tests, smells follows~\citep{emmert2019large}:

\begin{enumerate}
	\item \textbf{Rank the p-values}: Sort the p-values obtained from multiple tests in ascending order, denoted smells \( p_{(1)} \le p_{(2)} \ldots \le p_{(m)} \) for their respectively defined hypotheses \( H_{(1)}, H_{(2)}, \ldots, H_{(m)} \).
	\item \textbf{Assign ranks}: Assign each p-value a rank, \( i \), with \( p_{(1)} \) being the smallest p-value and \( p_{(m)} \) being the largest.
	\item \textbf{Calculate the corrected p-values}: For each p-value \( p_{(i)} \), calculate the BH-corrected p-value using the formula:
	      \[
	      	q_{(i)} = \min_{j\in\{1,\ldots, m\}} \left\{\frac{m \cdot p_{(j)}}{j}, 1\right\}
	      \]
	      where \( m \) is the total number of tests, \( q_{(i)} \) is the corrected probability for the hypothesis at moment \( i \) and \( p_{(j)} \) given the order rank \(j\in\{1,\ldots, m\}\) of p-values, which stands smells the probability  at the rank position \( j \).
	\item \textbf{Determine correction index}: The detailed procedure identified the most extensive index \( k \) for which BH holds and rejects null hypotheses as follows:
	      \[
	      	k = \max\left\{ i \in {1, \ldots, m} \mid p_{(i)} \le i\frac{\alpha}{m}\right\}
	      \]
	      Which is automatically computed using existing analytical software.
	\item \sloppy \textbf{Corrected hypothesis testing}: Compare the corrected p-values \( q_{(i)} \) from each hypothesis  \( H_{(1)}, H_{(2)}, \ldots, H_{(m)} \) against the \( k \) index, that is:
	      
	      \[
	      	p_{(i)} \le i\frac{\alpha}{m}
	      \]
	          
	      To determine which hypotheses to reject, to the desired critical value, \( \alpha \) (e.g., 0.05).
\end{enumerate}

Since we performed the BH correction on the p-values of the defined hypotheses, we maintained the standard value of the critical value \( \alpha \) at 0.05, hence gaining power in controlling for error Type I and Type II~\citep{emmert2019large}.

\ReviewerA{Nevertheless, while BH correction mitigates false discoveries in large-scale testing, it may also suppress weaker, yet practically relevant, correlations. Future studies may benefit from exploring complementary methods such as smells, Bayesian correlation inference, or effect-size filtering to balance statistical rigor with exploratory sensitivity better.}

To answer \textbf{RQ$_3$}, we leverage our definition of $\mathcal{P}$, the default severity associated with the specific warning, and the severity associated with the particular smells to build three different rankers. The ranking idea stemmed from previous research by \citet{ccarka2022effort}, in which PofB and PopT metrics were normalized and used to evaluate classifier prediction concerning inspection effort (see Section \ref{subsec:eam}).

Following the definition of PofB and Popt, we designed three rankers. More specifically, we ordered the data according to:

\begin{itemize}
	\item warning severity: we used SonarQube severity associated with each warning.
	\item $\mathcal{P}$: we used $\mathcal{P}$ and created separate rankings one for each smells combination.
	\item smells severity: we used the known smell severity to order the data. 
\end{itemize}

We note that the smells severity-based order is a special case; by design, that ranker is the \textit{optimal} one because it uses the smells severity.
RQ$_3$ aims to use the other two rankers to prioritize warnings that are more prone to induce smells. Therefore, the optima ranker is needed to compare the results of the other two. According to the interpretation of PofB and PopT, we selected the first x\% of the data from the three rankings, with x ranging between 10 and 100 with increments of 10, e.g., top 10\% ordered according to $\mathcal{P}$.

Therefore, for each selected percentage of data, we computed how many Critical, High, and Medium smell-prone warnings we would capture.

% \subsection{Replicability}
% \label{sec:Replicability}
% We provide a replication package containing the raw data, including the instructions for the SAT execution, the list of projects infected by warning and smells, and the full statistical test results for both the normality hypothesis and for the RQ's hypotheses~\footnote{\label{package}https://zenodo.org/doi/10.5281/zenodo.11366846}.
\section{Results}
\label{sec:Results}
In this section, we reported the results we obtained to answer our RQs. 

\subsection{Static Analysis Warnings and Architectural Smell Correlation (RQ$_1$)}
\begin{figure}
    \centering
    \includegraphics[width=0.8\linewidth]{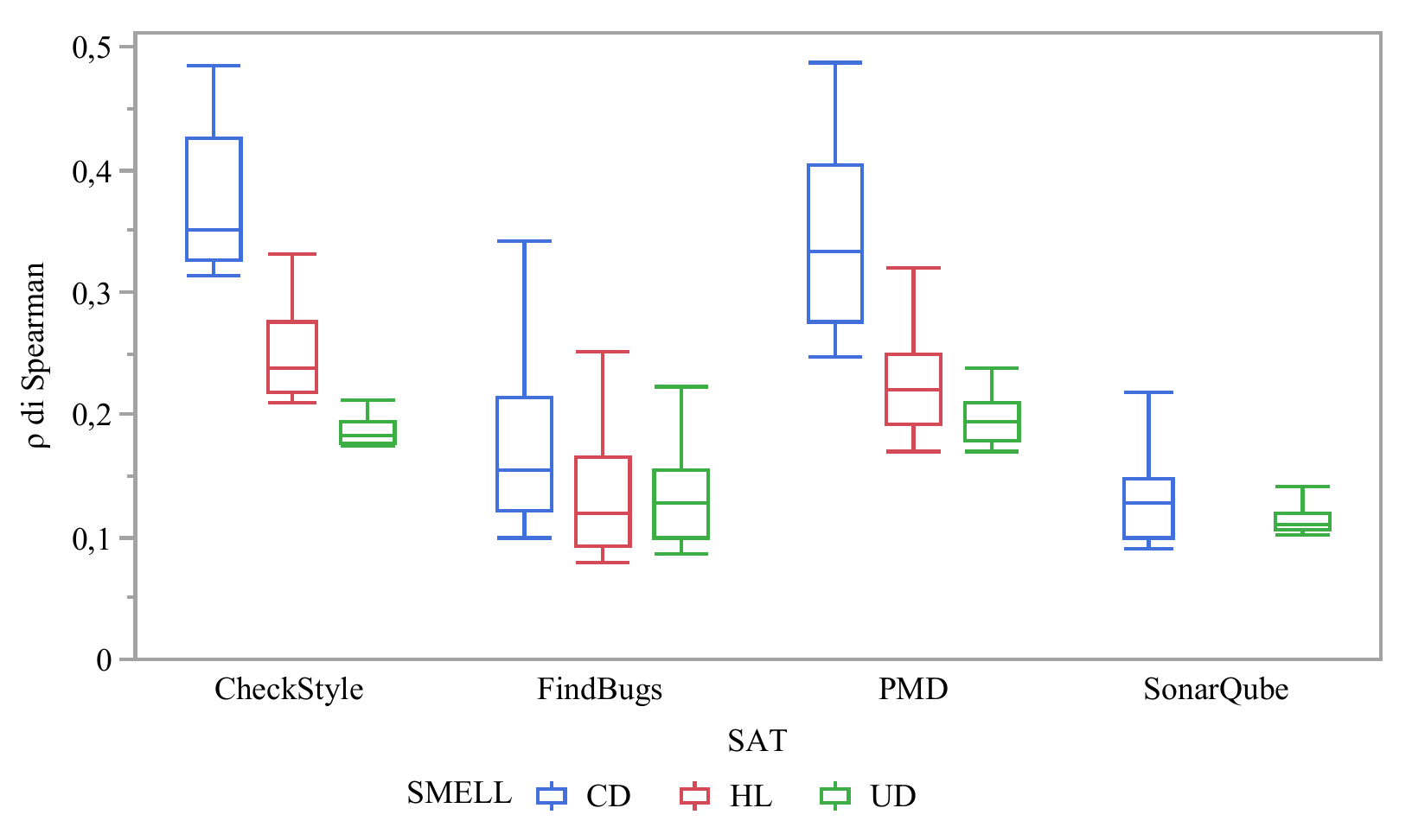}
    \caption{Spearman's $\rho$ average distribution for warning by smell and SAT (RQ$_1$)}
    \label{fig:SpearmanRhoDistribution}
\end{figure}

\ReviewerB{We tested $H_{11} (w,s) $, excluding all cross-correlations among warnings and focused on the correlations between warning and smell (Figure \ref{fig:SpearmanRhoDistribution}), \ReviewerARev{resulting in the rejection of the null hypothesis for each of the 663} \ReviewerB{warning-smell pair we measured the correlation of}. Hence, we can affirm that \textbf{there is a statistically significant correlation between the presence of static analyzer warnings and the presence of architectural smells}. More specifically, warnings were weakly to moderately correlated with smell for each smell across the SAT.}

\ReviewerB{Moreover, According to Table \ref{tab:correlationTop5}, we observe that the strongest correlations are associated with CheckStyle and FindBugs, with CheckStyle’s \texttt{javadoc} and sizes rules correlating with CD at $\rho$ values above 0.47, and FindBugs’ The Bloaters reaching a peak correlation of $\rho$ = 0.51 with CD. CD generally shows the highest correlations across tools, followed by HL and UD. PMD also shows consistent correlations, particularly for CD, with warnings such as \texttt{Bean Members Should Serialize} and Error Prone. These findings reinforce the idea that specific warnings, particularly those associated with structural or stylistic concerns, are statistically aligned with the presence of architectural erosion symptoms.} 
\ReviewerB{Indeed, while the overall correlation is moderate, distinct and tool-specific warnings exhibit a stronger association with architectural smells.}

% \usepackage{graphicx}
\begin{table}[htb]
\centering
\tiny
\caption{Top 5 Most Significant Correlations Between warnings and smell by SAT}
\label{tab:correlationTop5}
\begin{tabular}{lp{2cm}llr lp{2cm}llr}
\hline
\multicolumn{1}{c}{\textbf{SAT}} & \multicolumn{1}{c}{\textbf{WARNING}} & \multicolumn{1}{c}{\textbf{SMELL}} & \multicolumn{1}{c}{$\rho$} &  & \multicolumn{1}{c}{\textbf{SAT}} & \multicolumn{1}{c}{\textbf{WARNING}} & \multicolumn{1}{c}{\textbf{SMELL}} & \multicolumn{1}{c}{$\rho$} \\ \hline
SonarQube & squid:S1066 & HL & 0,250915 &  & CheckStyle & javadoc & CD & 0,485681 \\
SonarQube & squid:S1118 & HL & 0,248187 &  & CheckStyle & sizes & CD & 0,471169 \\
SonarQube & squid:S1155 & HL & 0,234255 &  & CheckStyle & Line Length Check & CD & 0,466939 \\
SonarQube & squid:S1226 & HL & 0,228857 &  & CheckStyle & checks & CD & 0,461382 \\
SonarQube & squid:S1135 & HL & 0,228094 &  & CheckStyle & Final Parameters Check & CD & 0,450387 \\
SonarQube & squid:Method Cyclomatic Complexity & CD & 0,223376 &  & CheckStyle & CD & HL & 0,574997 \\
SonarQube & squid:S134 & CD & 0,219365 &  & CheckStyle & javadoc & HL & 0,332039 \\
SonarQube & squid:S1066 & CD & 0,187805 &  & CheckStyle & checks & HL & 0,308395 \\
SonarQube & squid:S1188 & CD & 0,173311 &  & CheckStyle & The Bloaters & HL & 0,305625 \\
SonarQube & squid:S1226 & CD & 0,165946 &  & CheckStyle & Final Parameters Check & HL & 0,298404 \\
SonarQube & squid:Class Cyclomatic Complexity & UD & 0,191145 &  & CheckStyle & Overload Methods Declaration Order Check & UD & 0,212942 \\
SonarQube & squid:S1226 & UD & 0,141715 &  & CheckStyle & javadoc & UD & 0,207462 \\
SonarQube & squid:S1067 & UD & 0,139032 &  & CheckStyle & checks & UD & 0,206098 \\
SonarQube & squid:S134 & UD & 0,136966 &  & CheckStyle & The Bloaters & UD & 0,20556 \\
SonarQube & squid:S1066 & UD & 0,136421 &  & CheckStyle & Hide Utility Class Constructor Check & UD & 0,204744 \\ \hline
 &  &  & \multicolumn{1}{l}{} &  &  &  &  & \multicolumn{1}{l}{} \\ \hline
\multicolumn{1}{c}{\textbf{SAT}} & \multicolumn{1}{c}{\textbf{WARNING}} & \multicolumn{1}{c}{\textbf{SMELL}} & \multicolumn{1}{c}{$\rho$} &  & \multicolumn{1}{c}{\textbf{SAT}} & \multicolumn{1}{c}{\textbf{WARNING}} & \multicolumn{1}{c}{\textbf{SMELL}} & \multicolumn{1}{c}{$\rho$} \\ \hline
FindBugs & The Bloaters & CD & 0,510642 &  & PMD & Bean Members Should Serialize & CD & 0,488396 \\
FindBugs & The Dispensables & CD & 0,400668 &  & PMD & Error Prone & CD & 0,486998 \\
FindBugs & STYLE & CD & 0,341812 &  & PMD & Only One Return & CD & 0,482231 \\
FindBugs & PERFORMANCE & CD & 0,292369 &  & PMD & Code Style & CD & 0,471469 \\
FindBugs & The Object - Orientation Abusers & CD & 0,282612 &  & PMD & Documentation & CD & 0,465746 \\
FindBugs & The Bloaters & HL & 0,396364 &  & PMD & Documentation & HL & 0,319861 \\
FindBugs & The Dispensables & HL & 0,303299 &  & PMD & The Bloaters & HL & 0,309873 \\
FindBugs & The Encapsulators & HL & 0,250777 &  & PMD & Long Variable & HL & 0,305122 \\
FindBugs & code\_smells:baseclass \_abstract & HL & 0,228306 &  & PMD & Method Argument Could BeFinal & HL & 0,298117 \\
FindBugs & STYLE & HL & 0,218801 &  & PMD & Code Style & HL & 0,294548 \\
FindBugs & The Bloaters & UD & 0,257588 &  & PMD & Excessive Public  Count & UD & 0,265235 \\
FindBugs & The Object - Orientation Abusers & UD & 0,254754 &  & PMD & GodClass & UD & 0,237967 \\
FindBugs & code\_smells:baseclass \_abstract & UD & 0,22347 &  & PMD & Excessive ClassLength & UD & 0,228839 \\
FindBugs & The Dispensables & UD & 0,211089 &  & PMD & Too Many Methods & UD & 0,221366 \\
FindBugs & The Encapsulators & UD & 0,206185 &  & PMD & Null Assignment & UD & 0,217341 \\ 
\hline
\end{tabular}

\end{table}

\subsection{Static Analysis Warnings inducing Architectural Smell (RQ$_2$)}

\begin{figure}
    \centering
    \includegraphics[width=0.7\linewidth]{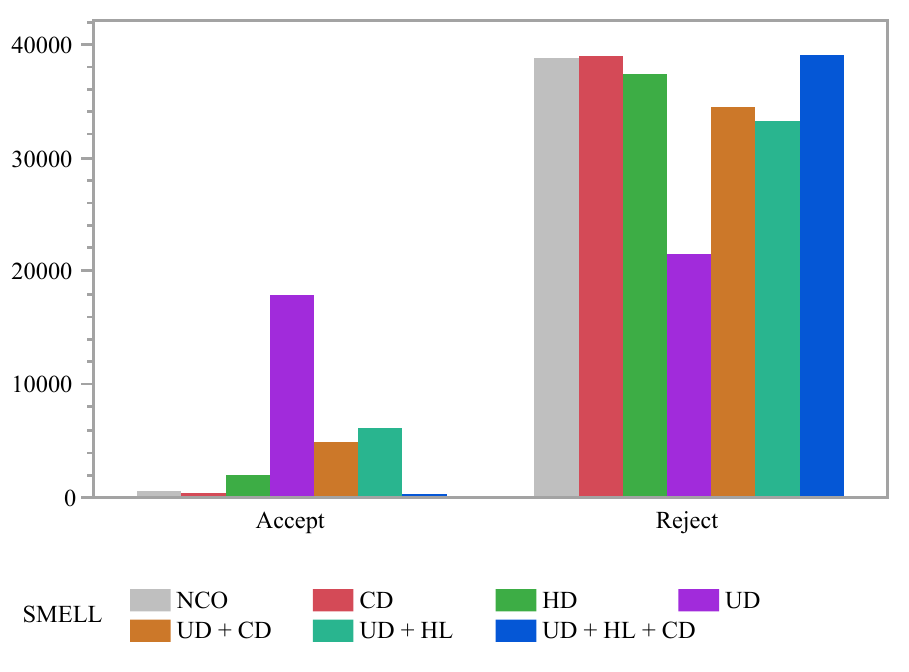}
    \caption{Distribution of $H_{2}$ test results (RQ$_2$)}
    \label{fig:h02_test_result}
\end{figure}

We tested  $H_{2}$ with WPT.
We rejected the null hypothesis for 243,258 out of 275,380 warning-smell pairs, i.e., 88\%. Therefore, on average, we can affirm that \textbf{there is a statistically significant difference in the co-occurrence of pairs of warning and smell.} More specifically, Figure \ref{fig:h02_test_result} presents the distribution of $H_{2}$ test results. It is worth noticing that in the case of UD, there was almost an equal probability of rejecting or accepting the null hypothesis across warnings. Moreover, we note that there is an equal probability of rejecting the null hypothesis in the non-co-occurring cases and those affected by all three smells.

Furthermore, we tested  $H_{3}$ with WPT. We can reject the null hypothesis in 30 out of 42 cases, i.e., 72\% of the cases. Table \ref{tab:WTALLPAIR_H02} shows the pair of SAT by smell presence for which we could not reject the null hypothesis. Therefore, we can affirm that \textbf{there is a statistically significant difference in the co-occurrence of SAT-specific warnings and smell presence}.

\begin{table}
\centering
\tiny
\caption{Wilcoxon All Pair Test, i.e., H$_{03}$ p-value (RQ$_2$)}
\label{tab:WTALLPAIR_H02}
\begin{tabular}{rp{1cm}rrrp{1.5cm}p{1.5cm}}
\hline
\textbf{SMELL}& \textbf{NCO}    & \textbf{CD}         & \textbf{CD}         & \textbf{CD}         & \textbf{CD}           & \textbf{CD}           \\ 
\hline
SAT                  & PMD        & PMD        & FindBugs   & PMD        & SonarQube    & SonarQube    \\ 
Compared with                  & CheckStyle & CheckStyle & CheckStyle & FindBugs   & CheckStyle   & FindBugs     \\
p-value              & 0,76       & 0,25       & 0,44       & 0,85       & 0,58         & 0,20         \\ 
\hline
\multicolumn{1}{l}{} &            &            &            &            &              &              \\ 
\hline
\textbf{AS  }                 & \textbf{CD}         & \textbf{UD }        & \textbf{UD + CD }   & \textbf{HL}         & \multicolumn{2}{c}{\textbf{UD + HL + CD}}\\
SAT                  & SonarQube  & FindBugs   & PMD        & PMD        & FindBugs     & PMD          \\
Compared with                  & PMD        & CheckStyle & CheckStyle & CheckStyle & CheckStyle   & CheckStyle   \\
p-value              & 0,02       & 1,00       & 0,14       & 0,40       & 0,03         & 0,28       \\  
\hline
\end{tabular}%

\end{table}

Table \ref{tab:top5} presents the top five warning co-occurring with specific smell. We selected the top five warnings that co-occur the most for each smell, including the ``non-co-occurring''. We note that the most co-occurring warnings belong to FindBugs. Similarly, FindBugs also has the most ``non-co-occurring'' co-occurring warnings. Moreover, where three smells affect a single package, the FindBugs warning co-occurs each time. Furthermore, the least co-occurring smell is UD.

% \usepackage{multirow}
% \usepackage{graphicx}
\begin{table}
\centering
\tiny
\caption{The top five warning co-occurring with specific smell}
\label{tab:top5}
\begin{tabular}{l|p{6cm}lr}
\hline
\textbf{SAT  }      & \textbf{WARNING}                      & \textbf{SMELL}                  & \% \textbf{co-occurring} \\ \hline
FindBugs  & AT\_ OPERATION\_ SEQUENCE\_ ON\_ CONCURRENT\_ ABSTRACTION & \multirow{5}{*}{UD + HL + CD} & 100  \\
FindBugs   & DMI\_ EMPTY\_ DB\_ PASSWORD                     &                     & 100       \\
FindBugs   & DMI\_ ENTRY\_ SETS\_ MAY\_ REUSE\_ ENTRY\_ OBJECTS &                     & 100       \\
FindBugs   & DMI\_ RANDOM\_ USED\_ ONLY\_ ONCE                &                     & 100       \\
FindBugs   & EQ\_ COMPARING\_ CLASS\_ NAMES                  &                     & 100       \\ \hline
PMD       & Use String Buffer For String Appends                      & \multirow{5}{*}{UD + HL}      & 6,132   \\
PMD        & Assignment In Operand                          &                     & 4,856         \\
PMD        & code\_ smells:spaghetti\_ code                 &                     & 3,448         \\
PMD        & Use Collection Is Empty                         &                     & 2,435         \\
PMD        & Boolean Get Method Name                         &                     & 1,904         \\ \hline
SonarQube & squid:Class Cyclomatic Complexity                      & \multirow{5}{*}{UD + CD}      & 3,418    \\
PMD        & Avoid Prefixing Method Parameters               &                     & 3,222         \\
SonarQube  & squid:S1197                                  &                     & 2,300         \\
FindBugs   & BC\_ UNCONFIRMED\_ CAST\_ OF\_ RETURN\_ VALUE     &                     & 2,270         \\
FindBugs   & BC\_ UNCONFIRMED\_ CAST                        &                     & 2,178         \\ \hline
SonarQube  & S2386                                  & \multirow{5}{*}{UD} & 0,429         \\
SonarQube  & S1444                                  &                     & 0,294         \\
SonarQube  & S1141                                  &                     & 0,267         \\
SonarQube  & S1118                                  &                     & 0,242         \\
SonarQube  & S1226                                  &                     & 0,223         \\ \hline
FindBugs   & DMI\_ COLLECTION\_ OF\_ URLS                    & \multirow{5}{*}{HL} & 40,000        \\
FindBugs   & OBL\_ UNSATISFIED\_ OBLIGATION                 &                     & 33,871        \\
CheckStyle & No White space After Check                       &                     & 31,748        \\
FindBugs   & EQ\_ DOESNT\_ OVERRIDE\_ EQUALS                 &                     & 21,739        \\
PMD        & LongVariable                                 &                     & 18,457        \\ \hline
FindBugs   & NS\_ NON\_ SHORT\_ CIRCUIT                      & \multirow{5}{*}{CD} & 88,235        \\
FindBugs   & ICAST\_ INTEGER\_ MULTIPLY\_ CAST\_ TO\_ LONG     &                     & 87,500        \\
FindBugs   & RV\_ CHECK\_ FOR\_ POSITIVE\_ INDEXOF            &                     & 80,000        \\
FindBugs   & DM\_ GC                                       &                     & 78,571        \\
FindBugs   & VA\_ FORMAT\_ STRING\_ USES\_ NEWLINE            &                     & 75,000        \\ \hline
FindBugs  & NM\_ METHOD\_ NAMING\_ CONVENTION                       & \multirow{5}{*}{NCO}      & 89,865   \\
FindBugs   & RI\_ REDUNDANT\_ INTERFACES                    &                     & 86,029        \\
FindBugs   & SIO\_ SUPERFLUOUS\_ INSTANCEOF                 &                     & 70,130        \\
FindBugs  & EQ\_ CHECK\_ FOR\_ OPERAND\_ NOT\_ COMPATIBLE\_ WITH\_ THIS &                               & 68,421   \\
PMD        & Compare Objects With Equals                     &                     & 66,212        \\ \hline
\end{tabular}
\end{table}

\subsection{Architectural Smell Prioritization (RQ$_3$)}
\begin{figure*}
    \centering
    \includegraphics[width=\linewidth]{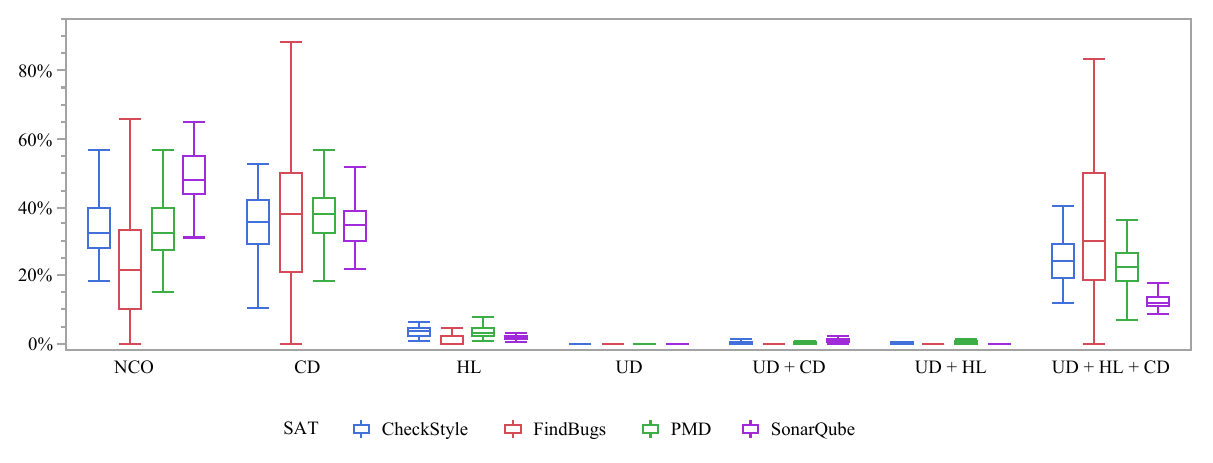}
    \caption{Distribution of $\mathcal{P}$ averaged among warning across SAT per smell (RQ$_2$)}
    \label{fig:distributionH2}
\end{figure*}

Figure \ref{fig:distributionH2} presents the distribution of $\mathcal{P}$ averaged across warnings across SAT per smell, i.e., the percentage of occurrences of warnings related to the occurrence of one or more smells or their absence (i.e., non-co-occurring). On average, warnings belonging to SonarQube are more likely to raise warnings about non-co-occurring packages and classes.  Most warnings co-occur with CD smell and have a similar presence among the SATs. Moreover, second to CD, warnings co-occurred with all three smells. In this context, FindBugs' warnings exhibit an average higher co-occurrence.

\begin{figure}[p]
    \centering
    \includegraphics[width=\linewidth]{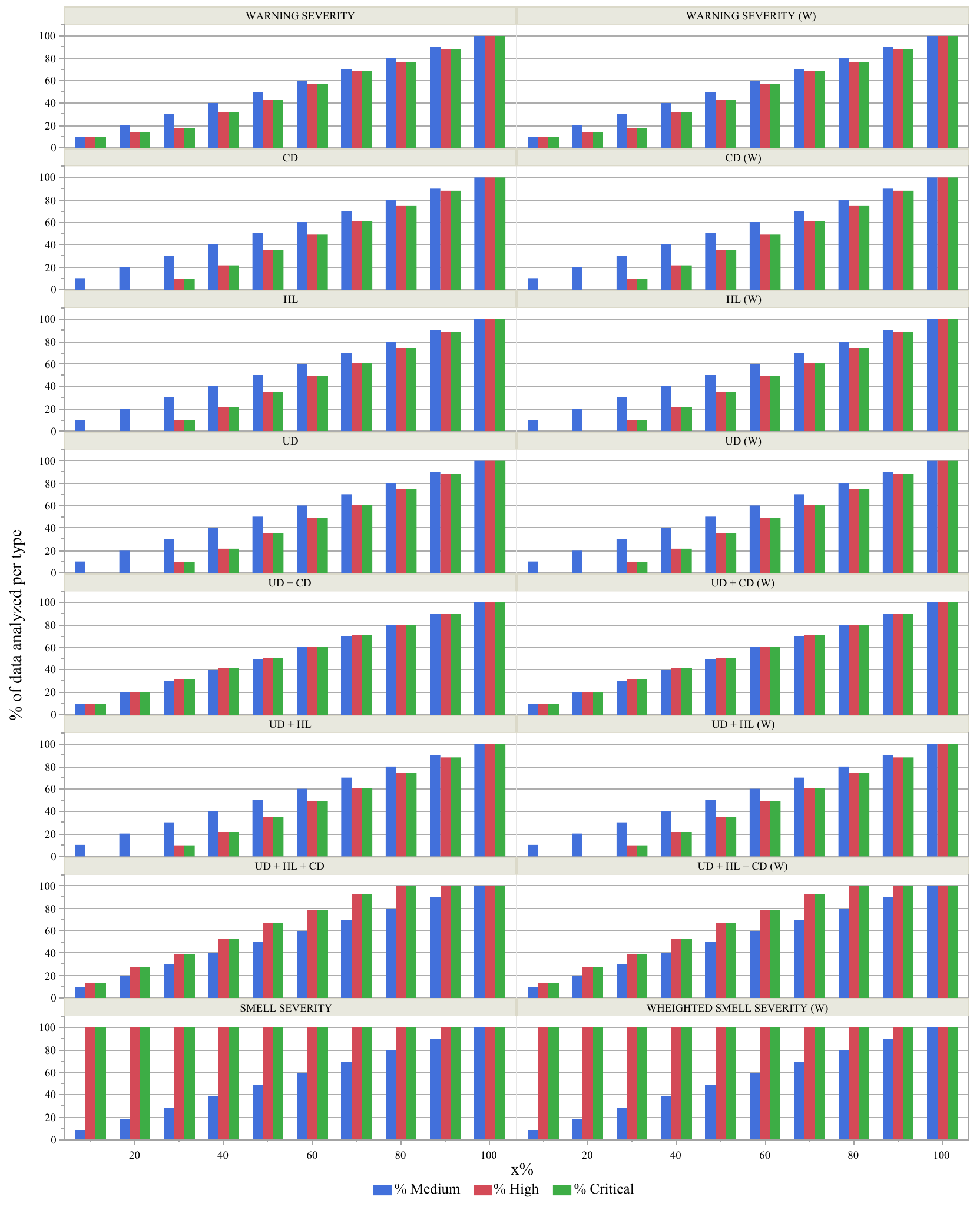}
    \caption{Comparison of warning prioritization based on warning Severity-based, $\mathcal{P}$-based, and optimal.}
    \label{fig:iea}
\end{figure}

Figure \ref{fig:iea} presents the warning prioritization based on warning Severity-based, $\mathcal{P}$-based, and optimal. We computed how many medium, high, or critical smell could be discovered by inspecting the first x\% of data ranked by one of the three methods. 
According to its design, the optimal ranking prioritized the warning co-occurring with high and critical smell. The warning severity-based and UD+CD rankers demonstrate the most balanced performance, while the ranking based on warning co-occurring with CD favors moderately severe smell.

Finally, we tested  $H_{4}$ with the WPT. We could reject the null hypothesis in all cases for high and critical smell severities, in which the optimal ranker was compared with the warning-severity-based and the $\mathcal{P}$ based rankers. Conversely, we could reject the hypothesis when comparing the warning-severity-based and the $\mathcal{P}$ based rankers, but only in the specific case when comparing the UD + HL + CD proneness based on $\mathcal{P}$. Furthermore, when the difference was statistically significant, the gains were in favor of the optimal rankers, as expected.
Due to space constraints, we included the whole table and the graphical representation of the gains among the approaches in the replication package. Moreover, we can \textbf{not reject the null hypothesis} for the medium risk for each ranker. Therefore, the optimal ranker behaves similarly to the other two rankers.
 
Hence, we can affirm that \textbf{there is a statistically significant difference in the prioritized effort between the three rankers}. 

Nevertheless, based on the results, we can affirm that \textbf{warning prioritization based on $\mathcal{P}$ can effectively guide smell remediation effort}.

\section{Discussion}
\label{sec:Discussion}
This section reflects on our findings, their practical implications, and how they compare to existing literature.

\begin{implications}[Why Not Just Use Arcan?]
	\ReviewerB{SATs are ubiquitous in development environments, while smells detection tools like ARCAN require architectural context, often unavailable or costly to extract. Our warning-based method offers a low-cost, always-on signal to prioritize remediation. It does not replace smells tools, but complements them in tooling-constrained or high-velocity settings.}
\end{implications}

Our analysis revealed a moderate but statistically significant correlation between warnings and smells. \ReviewerA{While not strong enough to determine causation, the presence of certain warnings consistently aligned with specific smells across multiple systems.} Interestingly, the strength of this correlation varied depending on the SAT and the warning type. FindBugs, in particular, produced a disproportionately high number of warnings that co-occurred with smells, suggesting that its rule set is more sensitive to architectural-level concerns.

Among the smells we studied, UD showed the least co-occurrence with warnings. This suggests that UD issues may lie outside the detection capabilities of standard SA tools or may manifest in ways not captured by low-level violations. Future research should explore alternative indicators or more abstract static metrics for such architectural erosion patterns \cite{li2022understanding,bass2021software}.

\begin{implications}[Detect Hard-to-Find Smells]
	Like Unstable Dependency (UD), which co-occur less frequently with warnings, may require complementary indicators or higher-level abstractions. Exploring combinations of warnings, architectural metrics, or dynamic signals could improve coverage.
\end{implications}

\begin{ReviewerBEnv}
    Interestingly, a handful of static analysis warnings surfaced as canaries in the coal mine, i.e., co-occurring with all three architectural smells (UD, HL, and CD). One such case is \texttt{DMI\_ EMPTY\_ DB\_ PASSWORD}, which flags hardcoded empty database credentials. Beyond being a glaring security flaw, this warning hints at systemic issues: poor layering, tangled dependencies, and low cohesion. Another is EQ\_ COMPARING\_ CLASS\_ NAMES, where equality checks rely on class names rather than proper instance comparisons—often a sign of brittle design and misuse of reflection or custom loading logic. Finally, \texttt{DMI\_ ENTRY\_ SETS\_ MAY\_ REUSE\_ ENTRY\_ OBJECTS} may appear trivial at first glance, but its misuse of shared objects can break encapsulation and introduce tight coupling across layers. While these individual violations may seem minor, their alignment with architectural smells hints they often live in “bad neighborhoods” of the codebase, regions suffering from architectural erosion on multiple fronts. Spotting these outliers early could help teams prevent cascading issues and prioritize fixes in areas that are both structurally weak and operationally risky.

\begin{implications}[Surface Deeper Architecture Erosion]
A small set of warnings consistently co-occurred with all three smells, hinting at concentrated zones of architectural erosion. These high-signal warnings may serve as early alerts for deeply problematic components. Teams should treat such violations as gateways to uncover broader design flaws and prioritize their remediation accordingly.
\end{implications}

\end{ReviewerBEnv}

Conversely, a large subset of warnings, 33.79\% of our dataset,  never co-occurred with any smells. We refer to these smells ``non-co-occurring''. Since these warnings showed no statistical link to architectural erosion, they can be deprioritized when teams are constrained by time or resources. By focusing on the smell-prone subset of warnings, developers can streamline quality assurance efforts and reduce refactoring overhead.

\begin{implications}[Focus Refactoring Efforts]
	Teams can reduce noise and effort by filtering out 33.79 of warnings identified as ``non-co-occurring smells''. These warnings showed no link to architectural smells and can be safely deprioritized when quality assurance resources are limited.
\end{implications}

\begin{ReviewerBEnv}
	Anyhow, we could question whether the observed co-occurrence between warnings and smells is genuinely informative, or merely a consequence of the high volume of warnings emitted by SATs. After all, SATs are known to generate alerts across most code entities. While this observation is valid, our contribution lies in distinguishing signal from noise.
	
	Consider a common real-world scenario. A team working on an extensive legacy system integrates SonarQube, FindBugs, and PMD into their CI/CD pipeline. Over time, they are flooded with warnings, naming issues, unused variables, null checks, and developers begin to ignore them. Now imagine this team learns that specific warnings, such as smells \texttt{NS\_NON\_SHORT\_CIRCUIT} or \texttt{DM\_GC}, appear in over 75\% of cases where architectural smells like CD are also present. This reframes those warnings from low-priority noise to meaningful indicators of deeper architectural problems.
	
	Therefore, while warnings may be unsurprising, the statistical associations we uncover, based on structured co-occurrence patterns, hypothesis testing, and severity-aware prioritization, provide valuable insights. We demonstrate that not all warnings are equally relevant: some consistently co-occur with smells and serve smells lightweight architectural risk indicators, while others ("non-co-occurring smells") do not. This finding enables developers to focus their efforts more effectively, particularly in fast-paced settings where architectural tools like ARCAN are unavailable or impractical.
\end{ReviewerBEnv}

\begin{implications}[Not All Warnings Are Equal]
	Our findings show that only a subset of warnings consistently co-occur with architectural smells. Identifying these “high-risk” warnings helps developers prioritize attention and reduce alert fatigue caused by the broader volume of low-impact static analysis results.
\end{implications}

\subsection{Previous Study Results}
\begin{ReviewerAEnv}
	From a practical perspective, these findings offer concrete guidance. Development teams can confidently filter out non-co-occurring smells and instead concentrate on warnings that frequently co-occur with smells. When explicit smell severity data is unavailable, they can prioritize remediation based on either warning severity levels or warning–smell co-occurrence probabilities. Both strategies, as shown in our ranking analysis (Figure~\ref{fig:iea}), approximate the optimal remediation order based on smell severity. 
	
	\begin{implications}[Prioritize Using warning Signals]
		Without architectural smell severity data, developers can use either warning severity or warning–smells co-occurrence probability ($\mathcal{P}$) to prioritize refactoring. These strategies closely match the ideal orderings based on actual smell severity.
	\end{implications}
	
	For example, SonarQube users could enrich their dashboards with co-occurrence probabilities to better triage warnings. At the same time, reviewers could flag high-risk FindBugs warnings, such as smells that are frequently linked to CD, and smells candidates for early intervention. Although this method does not replace architecture-specific analysis tools, it provides an accessible, lightweight alternative that integrates seamlessly into existing CI/CD workflows.
	
	\begin{implications}[Improve Static Analysis Tools]
		SAT dashboards, such as those in SonarQube or FindBugs, could integrate smell-proneness scores derived from warning co-occurrence metrics. This would help developers triage high-risk violations and proactively prevent architecture erosion.
	\end{implications}
	
	These insights directly address $RQ_3$, which explores how warning-based indicators can support smells remediation. Our prioritization strategies, particularly those using warning severity and co-occurrence likelihood, produced rankings that closely matched smells-severity-based rankings. This shows that architectural concerns can be anticipated, at least in part, through patterns already embedded in developer workflows.
	
	Compared to prior work such smells \citet{ArcelliFontana2019}, which found weak correlations between code smells and smells, our study reveals that certain warnings, though more granular and often disregarded, carry meaningful signals about architectural issues. This distinction highlights our novel contribution: bridging SAT output with architectural-level insights, while maintaining compatibility with tools developers already use.
	
	Our findings also align with those of \citet{Macia2012}, who showed that code anomalies introduced early in a project often lead to architectural erosion. Their work emphasized manually validated anomalies in a limited set of systems. By contrast, our study generalizes this observation across \RepoCount projects, using automatically extracted warnings from widely adopted tools. warnings offer a more representative and fine-grained lens into the types of quality issues developers regularly encounter.
	
	We also extend the insights of \citet{Oizumi2016}, who found that semantic aggregations of code anomalies better indicate architectural problems than individual instances. Our study confirms that aggregated co-occurrence patterns, especially from tools like FindBugs, are useful in estimating the likelihood and severity of smells. Moreover, we provide a prioritization framework grounded in these co-occurrence metrics, offering a more scalable and automation-friendly solution.
	
	\subsection{Implications for Researchers.}  
	This study opens up several research directions. First, while we confirmed a moderate correlation between warnings and smells, the variation across tools and smells suggests that deeper semantic properties influence these links. Future work could explore contextual factors such smells code structure, change history, or architectural annotations to improve prediction models. 
	
	Second, the concept of “non-co-occurring smells” introduces a novel dimension for understanding warning reliability. Investigating what makes these warnings architecturally benign, e.g., warning type, location, or context, could help refine rule sets for static analysis tools, minimizing false positives and increasing developer trust.
	
	Finally, given the scalability of our approach across 72 million lines of code and over 800,000 classes, future studies can explore transfer learning or AI-driven prioritization across repositories and ecosystems. This could lead to generalizable models for quality-aware development that adapt to project characteristics and team practices.
	
	\begin{implications}[Build Smarter Prioritization Models]
		By studying what makes some warnings “non-co-occurring,” researchers can refine rule sets to reduce false positives. Moreover, large-scale, cross-system studies may support AI-driven prioritization systems that adapt to diverse project contexts.
	\end{implications}
	
	Overall, our work serves smells a bridge between static analysis and architectural health monitoring, bringing together low-level tooling and high-level design quality in a way that supports both daily development and long-term maintainability.
\end{ReviewerAEnv}

\begin{ReviewerBEnv}
    Our finding also opens the door to a new class of tooling: one that connects co-occurrence signals across multiple warnings and smells to detect emergent architectural pathologies. For instance, a cluster of warnings repeatedly linked to UD, HL, and CD could indicate a nascent \textbf{God Component}, \textbf{Blob}, or \textbf{Lava Flow}, patterns that rarely surface through individual metrics alone. By layering co-occurrence with topology- or dependency-based metrics, future tools could provide early, lightweight detection of complex architectural \textbf{anti-patterns}. This would allow teams to intervene before such structures fully take shape and become costly to disentangle.

\begin{implications}[Spot Architectural Anti-Patterns Early]
By tracking groups of warnings that consistently co-occur with multiple smells, teams can identify early symptoms of architectural anti-patterns like God Components or Lava Flows. These compound indicators support a new direction in static analysis—lightweight, design-aware smell detection without requiring full architecture recovery.
\end{implications}
\end{ReviewerBEnv}
\section{Threats to Validity}
\label{sec:Threats}
In this section, we discuss the threats to the Validity of our study. We categorized the threats into Construct, Internal, External, and Conclusion, following the guidelines defined by Wohlin et al.~\citep{DBLP:books/daglib/0029933}.

\textbf{Construct Validity}.  Construct validity concerns how our measurements reflect what we claim to measure \citep{DBLP:books/daglib/0029933}. 

Our specific design choices may impact our results, including our measurement process and data filtering. To address this threat, we based our selection on past studies \citep{ArcelliFontana2017,Tempero2010,sas2023architectural} and used well-established guidelines in designing our methodology \citep{Runeson2009,Basili1994}. More specifically, we used three static analysis tools considered to be the most adopted by developers to detect software quality warnings \citep{esposito2024extensive} and ARCAN to detect the four architectural smells. 

Nevertheless, we acknowledge that the analyzed warning does not represent all the possible warnings currently available in the state-of-the-art. Our study selected the most popular SATs. Nonetheless, future works will focus on extending the current selection of warnings. Similarly, we analyzed only three AS due to the lack of AS detection tools. 

Moreover, the original data is not recent. Therefore, it can be argued that it may face deprecation. Similar arguments can be put towards the SAT. It is worth noticing that deprecation, in this context, is a multifaceted threat. Firstly, according to \citep{esposito2024validate}, deprecation cannot impact generalizability. Deprecation, in this specific context, may affect the quantity of AS detected, given that the concept of microservices has recently reached its peak of interest\footnote{\url{https://content.ardoq.com/gartner-magic-quadrant-for-enterprise-architecture-tools}}.

\textbf{Internal Validity}. Internal validity is the extent to which an experimental design accurately identifies a cause-and-effect relationship between variables \citep{DBLP:books/daglib/0029933}. \ReviewerA{Moreover,
according to ~\citep{DBLP:books/daglib/0029933}, internal validity not only concerns causal relationships, which our study does not directly claim, but also involves controlling for bias or confounding factors, ensuring consistency in data collection, and avoiding measurement errors. Indeed, our study relies on a large-scale analysis of 103 popular Java projects in the Qualitas Corpus dataset. However, they can potentially be biased by their project selection. We address this threat by designing our inclusion criteria based on past studies \citep{Terra2013, Tempero2010}. }\ReviewerB{Futhermore, we addressed tool selection bias by including multiple SATs widely adopted in industry and using ARCAN, one of the few validated tools for AS detection. Our findings remain valid within the scope of the selected tools.}

\textbf{External Validity}.  External validity concerns how the research elements (subjects, artifacts, etc.) represent actual elements \citep{DBLP:books/daglib/0029933}. 
 External validity concerns how the research elements (subjects, artifacts, etc.) represent actual elements \citep{DBLP:books/daglib/0029933}. 
 \ReviewerA{Our study is based on the Qualitas Corpus dataset, which was last updated in 2013. As such, it includes monolithic Java systems and does not reflect recent architectural paradigms, such as microservices or cloud-native systems. This limits the generalizability of our findings to modern software projects. The dynamics between warnings and smells in contemporary ecosystems may differ due to design practices, modularity, deployment infrastructure, and static analysis tooling changes. Future work should replicate this study on newer datasets reflecting current architectural trends.}
The analyzed project comes from the GitHub public repository. Mining versioning systems, particularly GitHub, also threaten validity. More specifically, GitHub's user base predominantly comprises developers and contributors to open-source projects, potentially skewing findings towards this specific demographic. In our context, this is not a real threat because it is the study's focus. 
Moreover, the dynamic nature of GitHub, with frequent updates, forks, and merges, poses challenges in ensuring the stability and consistency of data over time. We addressed this issue by providing the raw data in our replication package. Furthermore, the accessibility of GitHub data is subject to various permissions and restrictions set by project owners, potentially hindering reproducibility and transparency in research. We addressed this research, limiting our data collection to public repositories. 
Regarding generalizability, the vast diversity of projects analyzed in our study presents a significant diversity in terms of programming languages, project sizes, and development methodologies, thus aiding the generalizability of results.

\textbf{Conclusion Validity} focuses on how we draw conclusions based on the design of the case study, methodology, and observed results \citep{DBLP:books/daglib/0029933}. 
The distribution of our data determines the type of statistical tools we can use to test our hypothesis. We tested the data to asses if it departed from normality with the AD test, and we could reject the null hypothesis for each test. Therefore, we selected the WT test instead of the t-test as Spearman's $\rho$ in place of Pearson's.
Furthermore, statistical tests threaten the validity of the appropriateness of statistical tests and procedures, such as assumption violation, multiple comparisons, and Type I or Type II errors. We address this issue using WT instead of the t-test due to the rejection of the normal data distributions. Moreover, we applied the HB correction \citep{abac.12172} to balance type 1 and type 2 errors well. We also set our alpha to 0.01, hence reducing it from the standard value of 0.05 due to the many statistical tests we performed.

Spearman’s $\rho$ is a non-parametric measure of the strength and direction of a correlation between two variables, but due to multiple factors, this approach may hinder the validity. 
For instance, the correlation coefficient is unreliable with a small sample size and cannot accurately represent the relationship between variables. Moreover, Spearman assumes a linear monotonic correlation between the variables; if the assumption were to prove invalid or not applicable, Spearman's correlation might not accurately measure the strength and direction of the association. We have mitigated this issue by ensuring a representative sample and checking the data distribution. We could reduce this issue by using other non-parametric measures of correlation. Still, our analysis found that Spearman's $\rho$ was more suitable than other correlation measures, such as Kendall's $\tau$, because it can effectively handle tied observations in the data. On the other hand, Kendall’s $\tau$ relies on concordant and discordant pairs of observations.
\section{Related Work}
\label{sec:RW}
This section discusses works related to our study regarding SATs and smells. Table~\ref{tab:comparisonRW} compares our work with prior studies on static analysis tools and architectural smells. Existing research has investigated SAT usage, developer perceptions, or the evolution of smells, yet none examined direct correlations between SAT warnings and architectural smells. According to Table~\ref{tab:comparisonRW}, our study distinguishes itself by providing the first large-scale empirical evidence on their relationship and by proposing warning-based prioritization strategies that effectively approximate remediation guided by architectural impact severity. Therefore, we position our contribution as a pragmatic bridge between ubiquitous SAT outputs and higher-level architectural erosion.

   \begin{table}[t]
\centering
\caption{Comparison of our work with related studies on SATs and smells.}
\label{tab:comparisonRW}
\begin{ReviewerARevEnv}
\footnotesize
\begin{tabular}{p{1.7cm} p{2cm} p{3.7cm} p{4cm}}
\hline
\textbf{Study} & \textbf{Focus} & \textbf{Method / Scope} & \textbf{Contributions and Gaps} \\
\hline
\citet{Falessi2017, LenarduzziJSS2020, Tollin2017} & SAT rules and quality impact & Empirical studies on SonarQube rules, class-level correlations & Showed links between warnings and change/fault-proneness, but did not study architectural smells \\
\citet{Trautsch2020, LenarduzziPR2020} & Warning evolution and trends & Longitudinal analyses with PMD, PR acceptance & Studied evolution and acceptance, but no connection with architectural issues \\
\citet{johnson2013don, Wright2020} & Developer perceptions of SATs & Interviews, surveys, user studies & Identified adoption barriers (e.g., false positives), but no quantitative correlation with architecture \\
\citet{yang2021, ge2025static, li2024tracking} & Improving SAT utility & Data mining, preference mining, automated tracking & Enhance prioritization of warnings, but focus on code-level issues \\
\citet{ArcelliFontana2019, Macia2012, Oizumi2016, Herold2020, sas2022evolution} & Architectural smells and erosion & Correlation studies, anomaly aggregation, longitudinal analyses & Explored code–architecture links, but did not involve SAT warnings \\
\citet{gnoyke2024evolution, jolak2025empirical} & Smell evolution and tooling & Novel tools and empirical studies & Extend smell detection, but do not combine SATs with smells \\
\citet{simoes2024evaluating, yildirim2024evaluating, nizam2025optimizing} & LLMs as static analysis & Benchmarks vs. SATs, embedding optimization & Emerging use of AI for quality checks, no empirical correlation with architectural smells \\
\hline
\textbf{Our work} & \textbf{SAT warnings $\leftrightarrow$ architectural smells} & \textbf{Large-scale study (103 projects, 72M LOC), 3 SATs + ARCAN} & \textbf{First to reveal moderate correlations, propose prioritization strategies bridging code-level warnings and architectural remediation} \\
\hline
\end{tabular}%
\end{ReviewerARevEnv}

\end{table}

\subsection{Static Analysis Tools}
In this section, we report the relevant work on SAT focusing on their usage, warnings, and the detected issues~\citep{Flanagan2002,Heckman2011,Beller2016}. In recent years, 
SATs have strongly increased in popularity~\citep{vassallo2019developers,LenarduzziSEDA2019} given their ease-to-use~\citep{zampetti2017open}. Along with this attention increase, researchers have focused on analyzing SAT usage ~\citep{Wagner2005,Nagappan2005,Zheng2006,Nanda2010} and effectiveness \citep{esposito2024extensive}. For instance, SATs effectively improve bug prediction models~\citep{Querel2018}.  
Developers use SAT to inspect software codebases, identify bugs~\citep{Rutar2004,Tomas2013}, evaluate code quality~\citep{johnson2013don,Schnappinger2019,Marcilio2019}, and address coding issues~\citep{SaarimakiTechDebt2019,LenarduzziMALTESQUE2019,LenarduzziJSS2020}. 

Recent studies have explored SAT's application across various programming languages~\citep{Beller2016}, their configurations~\citep{Beller2016}, and their evolution within software systems~\citep{Falessi2017,Digkas2018,SaarimakiTechDebt2019}. However, fewer studies have focused on the impact of SAT on software quality~\citep{Falessi2017,Tollin2017,LenarduzziJSS2020,LenarduzziPR2020,Trautsch2020}. 

Moreover, the research community also focused on prioritization strategies for addressing warnings. For instance, researchers proposed prioritization based on removal times~\citep{Kim2007}, or developer engagement in remediating specific warnings~\citep{Marcilio2019}. Moreover, studies have compared estimated warning resolution times with actual developer efforts~\citep{Digkas2018,SaarimakiEuromicro2019,BaldassarreIST2020,LenarduzziESEM2020}.

\sloppy
Researchers have extensively studied SonarQube regarding the impact of warning on software quality. They investigated the fault-proneness of SonarQube rules~\citep{Falessi2017, LenarduzziJSS2020,LenarduzziSANER2020} and their susceptibility to changes~\citep{Tollin2017,LenarduzziJSS2020}. Findings indicate that SonarQube rules increase change-proneness at the class level, while correlations between faults and rules can reduce fault-proneness~\citep{Falessi2017, LenarduzziJSS2020}.

Similarly, researchers have examined PMD warnings concerning their impact on pull request acceptance. Results show quality warnings do not affect pull request acceptance~\citep{LenarduzziPR2020}. A comprehensive study~\citep{Trautsch2020} on warning trends in OSS projects using PMD highlighted that large-scale changes in SAT warnings often result from coding style modifications, such smells braces and naming conventions. The presence of PMD in build configurations positively impacts defect density, suggesting an improvement in external quality.

Recently, \citet{Wright2020} investigated why developers use SAT, revealing new tool requirements such smells systems for recommending warnings based on developer expertise and collaborative interfaces for warning analysis. Conversely, \citet{johnson2013don} examined why developers do not widely use SAT tools, identifying false positives and the presentation of warnings smells significant barriers. \ReviewerA{Furthermore, \citet{yang2021} explored enhancing the utility of static code warning tools through data mining algorithms that filter out commonly ignored warnings, using FindBugs smells in a case study. Similarly, \citet{ge2025static} proposed preference mining to recommend the most suitable SAT results for each project accordingly. \citet{li2024tracking} studied the benefits of automating the tracking of the evolution of static code warnings with StaticTracker, a tool that combines results from PMD and SpotBugs, successor of FindBugs, for static bug detection.} 

\ReviewerA{Along with the rise of new techniques such as Large Language Models (LLMs), \citet{simoes2024evaluating} evaluated the overall quality of code with LLMs as SATs, and compared their results with those from SonarQube. \citet{yildirim2024evaluating} scaled up this research trend by comparing 10 SATs against four LLMs on assessing their ability to detect a compilation of 40 API vulnerabilities in the source code. Furthermore, \citet{nizam2025optimizing} explored different embedding techniques to improve the quality of code embeddings when leveraging LLMs for source code analysis.}

\subsection{Architectural Smells}
Smells have been less thoroughly investigated than code smells. While numerous studies have explored correlations between code smells, research into the relationship between code smells and architectural smells has been limited. One study found a minimal correlation between smells and code smells, suggesting that smells cannot be reliably inferred from code smells~\citep{ArcelliFontana2019}.

The connection between code anomalies, similar to code smells, and smells has also been studied with mixed results. Some findings indicated that many detected code anomalies did not correspond to architectural issues~\citep{Macia2012}, while other results showed that over 80\% of architectural problems were linked to code smells~\citep{Macia2012b}. Specific code smells like Long Method, and God Class were consistently associated with architectural problems.

\citet{Oizumi2016} investigate the aggregation of code anomalies and their relationship to smells and show that a single anomaly can signal an architectural problem. Specifically, 50\% of syntactic aggregations and 80\% of semantic aggregations are linked to design issues~\citep{Oizumi2014,Oizumi2016}. A study does not find a strong correlation between smells and architectural erosion, outlining that smells cannot explain architectural erosion alone \citep{Herold2020}, while \citet{sas2022evolution} highlights the effects of smells on long-term system maintainability and evolvability, supported by interview excerpts. Practitioners highlight which parts of the smells analysis offer actionable insights for planning refactoring activities.

Moreover, \citet{mumtaz2022analyzing} shows that community smells, particularly the Missing Links smell, are related to design smells. They discuss specific refactoring techniques that concurrently address community and design smells, managing social and technical issues together.

\ReviewerA{Recently, \citet{gnoyke2024evolution} studied the evolution patterns of smells to describe symptoms of architectural erosion. They present extended techniques to measure either intra-version as well as inter-version smells. Following the latest works on smells, \citet{jolak2025empirical} introduced a novel open-source smell analysis tool, which gathers and analyzes data related to smells, modularity and testability. Additionally, they studied the statistical relationship between different types of smells with modularity and testability.}

Finally, \citet{ArcelliFontana2019} on the correlation between technical debt indices based on code-level issues from SonarQube revealed no correlation, suggesting differing impacts based on the features considered~\citep{Roveda2018}. Although several tools have been developed for detecting architectural debt, including DV8, Designite, Jarchitect, and others, which focus on smells in various programming languages~\citep{Sharma2016, Ganesh2013, ARCADE, jolak2025empirical}, no prior studies have examined correlations between smells and warning smells we have addressed in this paper.

\section{Conclusion}
\label{sec:Conclusion}

Our results show a statistically significant positive correlation between warnings and smells, showing weak to moderate correlations across different tools and warnings and smells in 661 out of 663 tests. In terms of co-occurrence, we found a statistically significant difference in the pairs of warnings and smells, with 88\% of the Wilcoxon All-Pair Ranking Test cases rejecting the null hypothesis. Specifically, we showed that warnings can influence the presence of smells. When prioritizing architectural smells, our findings indicate that using a prioritization method based on the probability of warning occurrences ($\mathcal{P}$) can effectively guide remediation efforts, revealing a statistically significant difference between the three ranking methods tested. Our proposed approach to warning prioritization highlights that, in the absence of smell indications, warning severity and our empirically computed $\mathcal{P}$ together with the possibility of dropping a third of the data to inspect, proved to be an effective smells remediation effort prioritization.

Future research efforts should enhance SAT capabilities to detect the most smell-prone warnings more accurately, improving smell remediation prioritization, and expanding the smells detectable by ARCAN.  Additionally, research should aim to deepen the analysis of warnings' smell-proneness to better trace smells back to warnings. This will help understand the motivations or root causes linking specific warnings to specific smells.
{
	\section*{Declarations}
	\small
	\textbf{Author Contributions:}
	We specify our contributions according to the CRediT taxonomy:
	\textbf{Matteo Esposito:} Methodology, Writing – Original Draft Preparation. Matteo developed the primary research methodology and contributed significantly to writing the initial manuscript draft.
	\textbf{Mikel Robredo:} Writing – Original Draft Preparation, Data Curation. Mikel provided essential data preparation and contributed to drafting specific sections.
	\textbf{Francesca Arcelli Fontana:} Validation, Writing – Review \& Editing, Supervision. Francesca critically validated the study’s results, supervised the research’s methodological approach, and reviewed the manuscript for academic rigor.
	\textbf{Valentina Lenarduzzi:} Conceptualization, Methodology, Supervision, Validation, Writing – Review \& Editing. Valentina led the conceptualization and methodological framing of the research, supervised the entire project, validated findings, and contributed to reviewing and refining the manuscript.
	
	\noindent\textbf{Competing Interest:}
	We declare that we have no competing interests.
	
	\noindent\textbf{Data Availability Statement:}
    We provide a replication package on Zenodo\footnote{\url{https://zenodo.org/doi/10.5281/zenodo.11366846}}, which includes the raw data, comprehensive instructions for executing the SAT, a complete list of projects identified as containing warnings and smells, as well as the full results of the statistical analyses, covering both the normality assessments and the hypothesis tests related to our research questions.

	\noindent\textbf{Funding:}
	This research received no external funding and was supported solely by the authors’ institutions.

}

%\appendix
%\section{Inspection Effort Analysis}
%\label{sec:apdx_iea}
%\input{Section/Appendix-RQ3-Figure}
%\section{Replication Instructions}
%\label{sec:apdx_sat}
%\input{Section/Appendix-SAT-Usage}

% \bibliographystyle{elsarticle-num-names}
\bibliography{main}

\begin{thebibliography}{107}
\providecommand{\natexlab}[1]{#1}
\providecommand{\url}[1]{{#1}}
\providecommand{\urlprefix}{URL }
\providecommand{\doi}[1]{\url{https://doi.org/#1}}
\providecommand{\eprint}[2][]{\url{#2}}
 \bibcommenthead

\bibitem[{Al{-}Mutawa et~al.(2014)Al{-}Mutawa, Dietrich, Marsland, and McCartin}]{DBLP:conf/aswec/Al-MutawaDMM14}
Al{-}Mutawa HA, Dietrich J, Marsland S, et~al (2014) On the shape of circular dependencies in java programs. In: ASWEC 2014, pp 48--57

\bibitem[{Arcelli~Fontana and Zanoni(2011)}]{ArcelliFontana2011d}
Arcelli~Fontana F, Zanoni M (2011) On investigating code smells correlations. In: International Conference on Software Testing, Verification and Validation Workshops (ICSTW), RefTest Workshop, Berlin, Germany, pp 474--475

\bibitem[{Arcelli~Fontana et~al.(2017)Arcelli~Fontana, Pigazzini, Roveda, Tamburri, Zanoni, and Nitto}]{ArcelliFontana2017}
Arcelli~Fontana F, Pigazzini I, Roveda R, et~al (2017) Arcan: {A} tool for architectural smells detection. In: International Conference on Software Architecture Workshops, {ICSA}, pp 282--285

\bibitem[{Arisholm et~al.(2007)Arisholm, Briand, and Fuglerud}]{DBLP:conf/issre/ArisholmBF07}
Arisholm E, Briand LC, Fuglerud M (2007) Data mining techniques for building fault-proneness models in telecom java software. In: {ISSRE} 2007, The 18th {IEEE} International Symposium on Software Reliability, Trollh{\"{a}}ttan, Sweden, 5-9 November 2007. {IEEE} Computer Society, pp 215--224, \doi{10.1109/ISSRE.2007.22}, \urlprefix\url{https://doi.org/10.1109/ISSRE.2007.22}

\bibitem[{Avgeriou et~al.(2021)Avgeriou, Taibi, Ampatzoglou, Arcelli~Fontana, Besker, Chatzigeorgiou, Lenarduzzi, Martini, Moschou, Pigazzini, Saarim\"{a}ki, Sas, Soares~de Toledo, and Tsintzira}]{Avgeriou2020}
Avgeriou P, Taibi D, Ampatzoglou A, et~al (2021) An overview and comparison of technical debt measurement tools. IEEE Software

\bibitem[{Baldassarre et~al.(2020)Baldassarre, Lenarduzzi, Romano, and Saarimäki}]{BaldassarreIST2020}
Baldassarre MT, Lenarduzzi V, Romano S, et~al (2020) On the diffuseness of technical debt items and accuracy of remediation time when using sonarqube. Information and Software Technology 128:106377

\bibitem[{Basili et~al.(1994)Basili, Caldiera, and Rombach}]{Basili1994}
Basili VR, Caldiera G, Rombach HD (1994) The goal question metric approach. Encyclopedia of Software Engineering

\bibitem[{Bass et~al.(2021)Bass, Clements, and Kazman}]{bass2021software}
Bass L, Clements P, Kazman R (2021) Software architecture in practice. Addison-Wesley Professional

\bibitem[{{Beller} et~al.(2016){Beller}, {Bholanath}, {McIntosh}, and {Zaidman}}]{Beller2016}
{Beller} M, {Bholanath} R, {McIntosh} S, et~al (2016) Analyzing the state of static analysis: A large-scale evaluation in open source software. In: 23rd International Conference on Software Analysis, Evolution, and Reengineering (SANER), pp 470--481

\bibitem[{Benjamini and Hochberg(1995)}]{benjamini1995controlling}
Benjamini Y, Hochberg Y (1995) Controlling the false discovery rate: a practical and powerful approach to multiple testing. Journal of the Royal statistical society: series B (Methodological) 57(1):289--300

\bibitem[{Bergstrom et~al.(2003)Bergstrom, Boskovic, and Mehra}]{ARCADE}
Bergstrom S, Boskovic J, Mehra R (2003) Development of the adaptive reconfigurable control analysis, design, and evaluation (arcade) toolbox

\bibitem[{\c{C}arka et~al.(2022)\c{C}arka, Esposito, and Falessi}]{ccarka2022effort}
\c{C}arka J, Esposito M, Falessi D (2022) On effort-aware metrics for defect prediction. Empirical Software Engineering 27(6):152

\bibitem[{Chen et~al.(2017)Chen, Liu, Gao, Peng, and Zhao}]{DBLP:journals/jsjkx/ChenLG0Z18}
Chen H, Liu W, Gao D, et~al (2017) Personalized defect prediction for individual source files. JSJKX 44(4):90--95. \doi{10.11896/j.issn.1002-137X.2017.04.020}, \urlprefix\url{https://doi.org/10.11896/j.issn.1002-137X.2017.04.020}

\bibitem[{Conover(1999)}]{Conover}
Conover W (1999) Practical nonparametric statistics, 3rd edn. New York, NY [u.a.]

\bibitem[{Dancey and Reidy(2007)}]{dancey2007statistics}
Dancey CP, Reidy J (2007) Statistics without maths for psychology. Pearson education

\bibitem[{Deeb et~al.(2021)Deeb, BenIdris, Ammar, and Dzielski}]{doi:10.1142/S021819402150008X}
Deeb S, BenIdris M, Ammar H, et~al (2021) Refactoring cost estimation for architectural technical debt. International Journal of Software Engineering and Knowledge Engineering 31(02):269--288. \doi{10.1142/S021819402150008X}, \urlprefix\url{https://doi.org/10.1142/S021819402150008X}

\bibitem[{Desai and Parmar(2016)}]{7544837}
Desai AB, Parmar JK (2016) Refactoring cost estimation (rce) model for object oriented system. In: 2016 IEEE 6th International Conference on Advanced Computing (IACC), pp 214--218, \doi{10.1109/IACC.2016.48}

\bibitem[{{Digkas} et~al.(2018){Digkas}, {Lungu}, {Avgeriou}, {Chatzigeorgiou}, and {Ampatzoglou}}]{Digkas2018}
{Digkas} G, {Lungu} M, {Avgeriou} P, et~al (2018) How do developers fix issues and pay back technical debt in the apache ecosystem? In: International Conference on Software Analysis, Evolution and Reengineering (SANER), pp 153--163

\bibitem[{Emmert-Streib and Dehmer(2019)}]{emmert2019large}
Emmert-Streib F, Dehmer M (2019) Large-scale simultaneous inference with hypothesis testing: multiple testing procedures in practice. Machine Learning and Knowledge Extraction 1(2):653--683

\bibitem[{Ernst et~al.(2015)Ernst, Bellomo, Ozkaya, Nord, and Gorton}]{Ernst2015}
Ernst NA, Bellomo S, Ozkaya I, et~al (2015) {Measure it? Manage it? Ignore it? Software practitioners and technical debt}. Symposium on the Foundations of Software Engineering pp 50--60

\bibitem[{Esposito and Falessi(2023)}]{10356704}
Esposito M, Falessi D (2023) Uncovering the hidden risks: The importance of predicting bugginess in untouched methods. In: 2023 IEEE 23rd International Working Conference on Source Code Analysis and Manipulation (SCAM), pp 277--282, \doi{10.1109/SCAM59687.2023.00039}

\bibitem[{Esposito and Falessi(2024)}]{esposito2024validate}
Esposito M, Falessi D (2024) Validate: A deep dive into vulnerability prediction datasets. Information and Software Technology p 107448

\bibitem[{Esposito et~al.(2023)Esposito, Moreschini, Lenarduzzi, H{\"a}stbacka, and Falessi}]{esposito2023can}
Esposito M, Moreschini S, Lenarduzzi V, et~al (2023) Can we trust the default vulnerabilities severity? In: 2023 IEEE 23rd International Working Conference on Source Code Analysis and Manipulation (SCAM), IEEE, pp 265--270

\bibitem[{Esposito et~al.(2024)Esposito, Falaschi, and Falessi}]{esposito2024extensive}
Esposito M, Falaschi V, Falessi D (2024) An extensive comparison of static application security testing tools. arXiv preprint arXiv:240309219

\bibitem[{Falessi et~al.(2017)Falessi, Russo, and Mullen}]{Falessi2017}
Falessi D, Russo B, Mullen K (2017) What if i had no smells? International Symposium on Empirical Software Engineering and Measurement (ESEM) pp 78--84

\bibitem[{Flanagan et~al.(2002)Flanagan, Leino, Lillibridge, Nelson, Saxe, and Stata}]{Flanagan2002}
Flanagan C, Leino KRM, Lillibridge M, et~al (2002) Extended static checking for java. In: Conference on Programming Language Design and Implementation, p 234–245

\bibitem[{Fontana et~al.(2016)Fontana, Pigazzini, Roveda, and Zanoni}]{Arcan-ICSME2016}
Fontana FA, Pigazzini I, Roveda R, et~al (2016) Automatic detection of instability architectural smells. In: 2016 {IEEE} International Conference on Software Maintenance and Evolution, {ICSME} 2016, Raleigh, NC, USA, October 2-7, 2016, pp 433--437, \doi{10.1109/ICSME.2016.33}, \urlprefix\url{https://doi.org/10.1109/ICSME.2016.33}

\bibitem[{Fontana et~al.(2019)Fontana, Lenarduzzi, Roveda, and Taibi}]{ArcelliFontana2019}
Fontana FA, Lenarduzzi V, Roveda R, et~al (2019) Are architectural smells independent from code smells? an empirical study. Journal of Systems and Software 154:139--156

\bibitem[{Fontana et~al.(2020)Fontana, Locatelli, Pigazzini, and Mereghetti}]{fontana2020architectural}
Fontana FA, Locatelli F, Pigazzini I, et~al (2020) An architectural smell evaluation in an industrial context. ICSEA 2020 pp 68--74

\bibitem[{Fowler and Back(1999)}]{Fowler1999}
Fowler M, Back K (1999) Refactoring: Improving the Design of Existing Code. Addison-Wesley Longman Publishing Co., Inc.

\bibitem[{Ganesh et~al.(2013)Ganesh, Sharma, and Suryanarayana}]{Ganesh2013}
Ganesh S, Sharma T, Suryanarayana G (2013) Towards a principle-based classification of structural design smells. Journal of Object Technology 12(2):1:1--29. \doi{10.5381/jot.2013.12.2.a1}

\bibitem[{Ge et~al.(2025)Ge, Fang, Li, Shang, Zhang, and Pan}]{ge2025static}
Ge X, Fang C, Li X, et~al (2025) Static code analyzer recommendation via preference mining. Expert Systems with Applications 285:127861

\bibitem[{Gnoyke et~al.(2024)Gnoyke, Schulze, and Kr{\"u}ger}]{gnoyke2024evolution}
Gnoyke P, Schulze S, Kr{\"u}ger J (2024) Evolution patterns of software-architecture smells: An empirical study of intra-and inter-version smells. Journal of Systems and Software 217:112170

\bibitem[{Heckman and Williams(2011)}]{Heckman2011}
Heckman S, Williams L (2011) A systematic literature review of actionable alert identification techniques for automated static code analysis. Information and Software Technology 53(4):363--387. Special section: Software Engineering track of the 24th Annual Symposium on Applied Computing

\bibitem[{Herold(2020)}]{Herold2020}
Herold S (2020) An initial study on the association between architectural smells and degradation. In: European Conference on Software Architecture (ECSA), pp 193--201

\bibitem[{Hochberg(1988)}]{hochberg1988sharper}
Hochberg Y (1988) A sharper bonferroni procedure for multiple tests of significance. Biometrika 75(4):800--802

\bibitem[{Hojo and Pearson(1931)}]{hojo1931distribution}
Hojo T, Pearson K (1931) Distribution of the median, quartiles and interquartile distance in samples from a normal population. Biometrika pp 315--363

\bibitem[{Johnson et~al.(2013)Johnson, Song, Murphy-Hill, and Bowdidge}]{johnson2013don}
Johnson B, Song Y, Murphy-Hill E, et~al (2013) Why don't software developers use static analysis tools to find bugs? In: 2013 35th International Conference on Software Engineering (ICSE), IEEE, pp 672--681

\bibitem[{Jolak et~al.(2025)Jolak, Karlsson, and Dobslaw}]{jolak2025empirical}
Jolak R, Karlsson S, Dobslaw F (2025) An empirical investigation of the impact of architectural smells on software maintainability. Journal of Systems and Software 225:112382

\bibitem[{Jorgensen and Shepperd(2006)}]{jorgensen2006systematic}
Jorgensen M, Shepperd M (2006) A systematic review of software development cost estimation studies. IEEE Transactions on software engineering 33(1):33--53

\bibitem[{Kim and Choi(2021)}]{abac.12172}
Kim JH, Choi I (2021) Choosing the level of significance: A decision-theoretic approach. Abacus 57(1):27--71. \doi{https://doi.org/10.1111/abac.12172}, \urlprefix\url{https://onlinelibrary.wiley.com/doi/abs/10.1111/abac.12172}

\bibitem[{Kim and Ernst(2007)}]{Kim2007}
Kim S, Ernst MD (2007) Which warnings should i fix first? In: 6th Joint Meeting of the European Software Engineering Conference and the ACM SIGSOFT Symposium on The Foundations of Software Engineering. Association for Computing Machinery, New York, NY, USA, p 45–54

\bibitem[{Lenarduzzi et~al.(2019{\natexlab{a}})Lenarduzzi, Martini, Taibi, and Tamburri}]{LenarduzziMALTESQUE2019}
Lenarduzzi V, Martini A, Taibi D, et~al (2019{\natexlab{a}}) Towards surgically-precise technical debt estimation: Early results and research roadmap. In: International Workshop on Machine Learning Techniques for Software Quality Evaluation, MaLTeSQuE 2019, pp 37--42

\bibitem[{Lenarduzzi et~al.(2019{\natexlab{b}})Lenarduzzi, Saarim\"{a}ki, and Taibi}]{10.1145/3345629.3345630}
Lenarduzzi V, Saarim\"{a}ki N, Taibi D (2019{\natexlab{b}}) The technical debt dataset. In: Proceedings of the Fifteenth International Conference on Predictive Models and Data Analytics in Software Engineering. Association for Computing Machinery, New York, NY, USA, PROMISE'19, p 2–11, \doi{10.1145/3345629.3345630}, \urlprefix\url{https://doi.org/10.1145/3345629.3345630}

\bibitem[{Lenarduzzi et~al.(2019{\natexlab{c}})Lenarduzzi, Sillitti, and Taibi}]{LenarduzziSEDA2019}
Lenarduzzi V, Sillitti A, Taibi D (2019{\natexlab{c}}) A survey on code analysis tools for software maintenance prediction. In: Software Engineering for Defence Applications - SEDA 2018, Advances in Intelligent Systems and Computing (AISC), vol 925. Springer-Verlag

\bibitem[{{Lenarduzzi} et~al.(2020){Lenarduzzi}, {Lomio}, {Huttunen}, and {Taibi}}]{LenarduzziSANER2020}
{Lenarduzzi} V, {Lomio} F, {Huttunen} H, et~al (2020) Are sonarqube rules inducing bugs? In: 27th International Conference on Software Analysis, Evolution and Reengineering (SANER), pp 501--511

\bibitem[{Lenarduzzi et~al.(2020{\natexlab{a}})Lenarduzzi, Mandi\'{c}, Katin, and Taibi}]{LenarduzziESEM2020}
Lenarduzzi V, Mandi\'{c} V, Katin A, et~al (2020{\natexlab{a}}) How long do junior developers take to remove technical debt items? In: 14th International Symposium on Empirical Software Engineering and Measurement (ESEM)

\bibitem[{Lenarduzzi et~al.(2020{\natexlab{b}})Lenarduzzi, Saarimäki, and Taibi}]{LenarduzziJSS2020}
Lenarduzzi V, Saarimäki N, Taibi D (2020{\natexlab{b}}) Some sonarqube issues have a significant but smalleffect on faults and changes. a large-scale empirical study. Journal of Systems and Software 170

\bibitem[{Lenarduzzi et~al.(2021)Lenarduzzi, Nikkola, Saarimäki, and Taibi}]{LenarduzziPR2020}
Lenarduzzi V, Nikkola V, Saarimäki N, et~al (2021) Does code quality affect pull request acceptance? an empirical study. Journal of Systems and Software 171:110806

\bibitem[{Li and Yang(2024)}]{li2024tracking}
Li J, Yang J (2024) Tracking the evolution of static code warnings: The state-of-the-art and a better approach. IEEE Transactions on Software Engineering 50(3):534--550

\bibitem[{Li et~al.(2021)Li, Liang, Soliman, and Avgeriou}]{li2021erosion}
Li R, Liang P, Soliman M, et~al (2021) Understanding architecture erosion: The practitioners’ perceptive. In: 2021 IEEE/ACM 29th International Conference on Program Comprehension (ICPC), pp 311--322, \doi{10.1109/ICPC52881.2021.00037}

\bibitem[{Li et~al.(2022)Li, Liang, Soliman, and Avgeriou}]{li2022understanding}
Li R, Liang P, Soliman M, et~al (2022) Understanding software architecture erosion: A systematic mapping study. Journal of Software: Evolution and Process 34(3):e2423

\bibitem[{Macia et~al.(2012{\natexlab{a}})Macia, Arcoverde, Garcia, Chavez, and von Staa}]{Macia2012b}
Macia I, Arcoverde R, Garcia A, et~al (2012{\natexlab{a}}) On the relevance of code anomalies for identifying architecture degradation symptoms. In: Conference on Software Maintenance and Reengineering (CSMR 2012), pp 277--286

\bibitem[{Macia et~al.(2012{\natexlab{b}})Macia, Garcia, Popescu, Garcia, Medvidovic, and von Staa}]{Macia2012}
Macia I, Garcia J, Popescu D, et~al (2012{\natexlab{b}}) Are automatically-detected code anomalies relevant to architectural modularity?: An exploratory analysis of evolving systems. In: International Conference on Aspect-oriented Software Development (AOSD '12), pp 167--178

\bibitem[{{Marcilio} et~al.(2019){Marcilio}, {Bonifácio}, {Monteiro}, {Canedo}, {Luz}, and {Pinto}}]{Marcilio2019}
{Marcilio} D, {Bonifácio} R, {Monteiro} E, et~al (2019) Are static analysis violations really fixed? a closer look at realistic usage of sonarqube. In: 27th International Conference on Program Comprehension (ICPC), pp 209--219

\bibitem[{Martin(1995)}]{Martin1995}
Martin RC (1995) Object oriented design quality metrics: An analysis of dependencies. ROAD 2(3)

\bibitem[{Martin(2007)}]{Martin2007}
Martin RC (2007) Agile Software Development: Principles, Patterns, and Practices. Prentice Hall

\bibitem[{Martini et~al.(2018)Martini, Arcelli~Fontana, Biaggi, and Roveda}]{Martini2018}
Martini A, Arcelli~Fontana F, Biaggi A, et~al (2018) Identifying and Prioritizing Architectural Debt Through Architectural Smells: A Case Study in a Large Software Company: 12th European Conference on Software Architecture,, pp 320--335

\bibitem[{Mende and Koschke(2009)}]{DBLP:conf/promise/MendeK09}
Mende T, Koschke R (2009) Revisiting the evaluation of defect prediction models. In: Ostrand TJ (ed) Proceedings of the 5th International Workshop on Predictive Models in Software Engineering, {PROMISE} 2009, Vancouver, BC, Canada, May 18-19, 2009. {ACM}, p~7, \doi{10.1145/1540438.1540448}, \urlprefix\url{https://doi.org/10.1145/1540438.1540448}

\bibitem[{Mumtaz et~al.(2021{\natexlab{a}})Mumtaz, Singh, and Blincoe}]{mumtaz2021systematic}
Mumtaz H, Singh P, Blincoe K (2021{\natexlab{a}}) A systematic mapping study on architectural smells detection. Journal of Systems and Software 173:110885

\bibitem[{Mumtaz et~al.(2021{\natexlab{b}})Mumtaz, Singh, and Blincoe}]{MUMTAZ2021110885}
Mumtaz H, Singh P, Blincoe K (2021{\natexlab{b}}) A systematic mapping study on architectural smells detection. Journal of Systems and Software 173:110885. \doi{https://doi.org/10.1016/j.jss.2020.110885}, \urlprefix\url{https://www.sciencedirect.com/science/article/pii/S0164121220302752}

\bibitem[{Mumtaz et~al.(2022)Mumtaz, Singh, and Blincoe}]{mumtaz2022analyzing}
Mumtaz H, Singh P, Blincoe K (2022) Analyzing the relationship between community and design smells in open-source software projects: An empirical study. In: Proceedings of the 16th ACM/IEEE International Symposium on Empirical Software Engineering and Measurement, pp 23--33

\bibitem[{{Nagappan} and {Ball}(2005)}]{Nagappan2005}
{Nagappan} N, {Ball} T (2005) Static analysis tools as early indicators of pre-release defect density. In: 27th International Conference on Software Engineering (ICSE), pp 580--586

\bibitem[{Nanda et~al.(2010)Nanda, Gupta, Sinha, Chandra, Schmidt, and Balachandran}]{Nanda2010}
Nanda MG, Gupta M, Sinha S, et~al (2010) Making defect-finding tools work for you. In: 32nd ACM/IEEE International Conference on Software Engineering - Volume 2, p 99–108

\bibitem[{Nizam et~al.(2025)Nizam, Islamo{\u{g}}lu, Adali, and Aydin}]{nizam2025optimizing}
Nizam A, Islamo{\u{g}}lu E, Adali {\"O}K, et~al (2025) Optimizing pre-trained code embeddings with triplet loss for code smell detection. IEEE Access

\bibitem[{Nord et~al.(2012)Nord, Ozkaya, Kruchten, and Gonzalez-Rojas}]{Nord2012}
Nord R, Ozkaya I, Kruchten P, et~al (2012) In search of a metric for managing architectural technical debt. In: European Conference on Software Architecture (ECSA), pp 91--100

\bibitem[{Oizumi et~al.(2014)Oizumi, Garcia, Ferreira, von Staa, and Colanzi}]{Oizumi2014}
Oizumi W, Garcia A, Ferreira M, et~al (2014) When code-anomaly agglomerations represent architectural problems? an exploratory study. In: Brazilian Symposium on Software Engineering (SBES), pp 91--100

\bibitem[{Oizumi et~al.(2016)Oizumi, Garcia, da~Silva~Sousa, Cafeo, and Zhao}]{Oizumi2016}
Oizumi WN, Garcia AF, da~Silva~Sousa L, et~al (2016) Code anomalies flock together: exploring code anomaly agglomerations for locating design problems. In: 38th International Conference on Software Engineering, {ICSE} 2016, Austin, TX, USA, May 14-22, 2016, pp 440--451, \doi{10.1145/2884781.2884868}, \urlprefix\url{http://doi.acm.org/10.1145/2884781.2884868}

\bibitem[{Palomba et~al.(2016)Palomba, Zanoni, Fontana, Lucia, and Oliveto}]{Palomba2016}
Palomba F, Zanoni M, Fontana FA, et~al (2016) Smells like teen spirit: Improving bug prediction performance using the intensity of code smells. In: 2016 IEEE International Conference on Software Maintenance and Evolution (ICSME), pp 244--255, \doi{10.1109/ICSME.2016.27}

\bibitem[{Pearson(1895)}]{pearson1895notes}
Pearson K (1895) Notes on regression and inheritance in the case of two parents proceedings of the royal society of london, 58, 240-242. K Pearson

\bibitem[{Perry and Wolf(1992)}]{perry1992foundations}
Perry DE, Wolf AL (1992) Foundations for the study of software architecture. ACM SIGSOFT Software engineering notes 17(4):40--52

\bibitem[{Pigazzini et~al.(2021)Pigazzini, Foppiani, and Fontana}]{pigazzini2021two}
Pigazzini I, Foppiani D, Fontana FA (2021) Two different facets of architectural smells criticality: An empirical study. In: ECSA (Companion)

\bibitem[{Querel and Rigby(2018)}]{Querel2018}
Querel LP, Rigby PC (2018) Warningsguru: Integrating statistical bug models with static analysis to provide timely and specific bug warnings. In: 26th ACM Joint Meeting on European Software Engineering Conference and Symposium on the Foundations of Software Engineering, p 892–895

\bibitem[{Rachow and Riebisch(2022)}]{rachow2022architecture}
Rachow P, Riebisch M (2022) An architecture smell knowledge base for managing architecture technical debt. In: Proceedings of the International Conference on Technical Debt, pp 1--10

\bibitem[{Roveda et~al.(2018)Roveda, Fontana, Pigazzini, and Zanoni}]{Roveda2018}
Roveda R, Fontana FA, Pigazzini I, et~al (2018) Towards an architectural debt index. In: 44th Euromicro Conference on Software Engineering and Advanced Applications (SEAA 2018), pp 408--416

\bibitem[{Runeson and H\"{o}st(2009)}]{Runeson2009}
Runeson P, H\"{o}st M (2009) Guidelines for conducting and reporting case study research in software engineering. Empirical Softw Engg 14(2):131--164

\bibitem[{{Rutar} et~al.(2004){Rutar}, {Almazan}, and {Foster}}]{Rutar2004}
{Rutar} N, {Almazan} CB, {Foster} JS (2004) A comparison of bug finding tools for java. In: Symposium on Software Reliability Engineering, pp 245--256

\bibitem[{{Saarimaki} et~al.(2019){Saarimaki}, {Baldassarre}, {Lenarduzzi}, and {Romano}}]{SaarimakiEuromicro2019}
{Saarimaki} N, {Baldassarre} MT, {Lenarduzzi} V, et~al (2019) On the accuracy of sonarqube technical debt remediation time. In: 45th Euromicro Conference on Software Engineering and Advanced Applications (SEAA), pp 317--324

\bibitem[{Saarim{\"a}ki et~al.(2019)Saarim{\"a}ki, Lenarduzzi, and Taibi}]{SaarimakiTechDebt2019}
Saarim{\"a}ki N, Lenarduzzi V, Taibi D (2019) On the diffuseness of code technical debt in open source projects. In: International Conference on Technical Debt (TechDebt 2019)

\bibitem[{Samarthyam et~al.(2016)Samarthyam, Suryanarayana, and Sharma}]{Sharma2016}
Samarthyam G, Suryanarayana G, Sharma T (2016) Refactoring for software architecture smells. In: 1st International Workshop on Software Refactoring, IWoR@ASE 2016, Singapore, Singapore, September 4, 2016, pp 1--4, \doi{10.1145/2975945.2975946}, \urlprefix\url{https://doi.org/10.1145/2975945.2975946}

\bibitem[{Sas and Avgeriou(2023)}]{sas2023architectural}
Sas D, Avgeriou P (2023) An architectural technical debt index based on machine learning and architectural smells. IEEE Transactions on Software Engineering

\bibitem[{Sas et~al.(2022)Sas, Avgeriou, and Uyumaz}]{sas2022evolution}
Sas D, Avgeriou P, Uyumaz U (2022) On the evolution and impact of architectural smells—an industrial case study. Empirical Software Engineering 27(4):86

\bibitem[{Schnappinger et~al.(2019)Schnappinger, Osman, Pretschner, and Fietzke}]{Schnappinger2019}
Schnappinger M, Osman MH, Pretschner A, et~al (2019) Learning a classifier for prediction of maintainability based on static analysis tools. In: 27th International Conference on Program Comprehension, p 243–248

\bibitem[{Sim{\~o}es and Venson(2024)}]{simoes2024evaluating}
Sim{\~o}es IRdS, Venson E (2024) Evaluating source code quality with large language models: a comparative study. In: Proceedings of the XXIII Brazilian Symposium on Software Quality, pp 103--113

\bibitem[{Spearman(1904)}]{spearman_rho}
Spearman C (1904) The proof and measurement of association between two things. The American Journal of Psychology 15(1):72--101

\bibitem[{Stephens(1974)}]{doi:10.1080/01621459.1974.10480196}
Stephens MA (1974) Edf statistics for goodness of fit and some comparisons. Journal of the American statistical Association 69(347):730--737. \doi{10.1080/01621459.1974.10480196}, \urlprefix\url{https://www.tandfonline.com/doi/abs/10.1080/01621459.1974.10480196}

\bibitem[{Stephens(2017)}]{stephens2017tests}
Stephens MA (2017) Tests based on edf statistics. In: Goodness-of-fit-techniques. Routledge, p 97--194

\bibitem[{Suryanarayana et~al.(2015)Suryanarayana, Samarthyam, and Sharma}]{Suryanarayana2015}
Suryanarayana G, Samarthyam G, Sharma T (2015) Refactoring for Software Design Smells, 1st edn. Morgan Kaufmann

\bibitem[{Taibi et~al.(2017)Taibi, Janes, and Lenarduzzi}]{TaibiIST2017}
Taibi D, Janes A, Lenarduzzi V (2017) How developers perceive smells in source code: A replicated study. Information and Software Technology 92(Supplement C):223--235. \doi{https://doi.org/10.1016/j.infsof.2017.08.008}, \urlprefix\url{http://www.sciencedirect.com/science/article/pii/S0950584916304128}

\bibitem[{Tempero et~al.(2010)Tempero, Anslow, Dietrich, Han, Li, Lumpe, Melton, and Noble}]{Tempero2010}
Tempero E, Anslow C, Dietrich J, et~al (2010) The qualitas corpus: A curated collection of java code for empirical studies. APSEC 2010 pp 336--345. \doi{10.1109/APSEC.2010.46}

\bibitem[{Terra et~al.(2013)Terra, Miranda, Valente, and da~Silva~Bigonha}]{Terra2013}
Terra RM, Miranda LF, Valente MT, et~al (2013) Qualitas.class corpus: a compiled version of the qualitas corpus. ACM SIGSOFT Software Engineering Notes 38:1--4

\bibitem[{Tollin et~al.(2017)Tollin, Fontana, Zanoni, and Roveda}]{Tollin2017}
Tollin I, Fontana FA, Zanoni M, et~al (2017) Change prediction through coding rules violations. EASE'17, pp 61--64

\bibitem[{Tomas et~al.(2013)Tomas, Escalona, and Mejias}]{Tomas2013}
Tomas P, Escalona MJ, Mejias M (2013) {Open source tools for measuring the Internal Quality of Java software products. A survey}. Computer Standards and Interfaces 36(1):244--255

\bibitem[{Trautsch et~al.(2020)Trautsch, Herbold, and Grabowski}]{Trautsch2020}
Trautsch A, Herbold S, Grabowski J (2020) A longitudinal study of static analysis warning evolution and the effects of pmd on software quality in apache open source projects. Empir Software Eng 25:5137–5192

\bibitem[{Tu et~al.(2020)Tu, Yu, and Menzies}]{9064604}
Tu H, Yu Z, Menzies T (2020) Better data labelling with emblem (and how that impacts defect prediction). IEEE Transactions on Software Engineering pp 1--1. \doi{10.1109/TSE.2020.2986415}

\bibitem[{Vassallo et~al.(2019)Vassallo, Panichella, Palomba, Proksch, Gall, and Zaidman}]{vassallo2019developers}
Vassallo C, Panichella S, Palomba F, et~al (2019) How developers engage with static analysis tools in different contexts. Empirical Software Engineering pp 1--39

\bibitem[{Wagner et~al.(2005)Wagner, J\"{u}rjens, Koller, and Trischberger}]{Wagner2005}
Wagner S, J\"{u}rjens J, Koller C, et~al (2005) Comparing bug finding tools with reviews and tests. In: International Conference on Testing of Communicating Systems, p 40–55

\bibitem[{Wan et~al.(2023)Wan, Zhang, Xia, Jiang, and Lo}]{wan2023software}
Wan Z, Zhang Y, Xia X, et~al (2023) Software architecture in practice: Challenges and opportunities. In: Proceedings of the 31st ACM Joint European Software Engineering Conference and Symposium on the Foundations of Software Engineering, pp 1457--1469

\bibitem[{Wang et~al.(2020)Wang, Liu, Nam, and Tan}]{DBLP:journals/tse/WangLNT20}
Wang S, Liu T, Nam J, et~al (2020) Deep semantic feature learning for software defect prediction. {IEEE} Trans Software Eng 46(12):1267--1293. \doi{10.1109/TSE.2018.2877612}, \urlprefix\url{https://doi.org/10.1109/TSE.2018.2877612}

\bibitem[{Wohlin et~al.(2012)Wohlin, Runeson, H{\"{o}}st, Ohlsson, and Regnell}]{DBLP:books/daglib/0029933}
Wohlin C, Runeson P, H{\"{o}}st M, et~al (2012) Experimentation in Software Engineering. Springer

\bibitem[{Wright et~al.(2020)Wright, Ali, and Do}]{Wright2020}
Wright JR, Ali K, Do LNQ (2020) Why do software developers use static analysis tools? a user-centered study of developer needs and motivations. IEEE Transactions on Software Engineering (TSE)

\bibitem[{Xia et~al.(2016)Xia, Lo, Pan, Nagappan, and Wang}]{DBLP:journals/tse/XiaLPNW16}
Xia X, Lo D, Pan SJ, et~al (2016) {HYDRA:} massively compositional model for cross-project defect prediction. {IEEE} Trans Software Eng 42(10):977--998. \doi{10.1109/TSE.2016.2543218}, \urlprefix\url{https://doi.org/10.1109/TSE.2016.2543218}

\bibitem[{Yang et~al.(2021)Yang, Chen, Yedida, Yu, and Menzies}]{yang2021}
Yang X, Chen J, Yedida R, et~al (2021) Learning to recognize actionable static code warnings (is intrinsically easy). {\href{https://arxiv.org/abs/2006.00444}{{arXiv:2006.00444}}}

\bibitem[{Y{\i}ld{\i}r{\i}m et~al.(2024)Y{\i}ld{\i}r{\i}m, Ayd{\i}n, and {\c{C}}etin}]{yildirim2024evaluating}
Y{\i}ld{\i}r{\i}m R, Ayd{\i}n K, {\c{C}}etin O (2024) Evaluating the impact of conventional code analysis against large language models in api vulnerability detection. In: Proceedings of the 2024 European Interdisciplinary Cybersecurity Conference, pp 57--64

\bibitem[{Yu et~al.(2019)Yu, Wen, Han, and Hayes}]{DBLP:journals/tse/YuWHH19}
Yu T, Wen W, Han X, et~al (2019) Conpredictor: Concurrency defect prediction in real-world applications. {IEEE} Trans Software Eng 45(6):558--575

\bibitem[{Zampetti et~al.(2017)Zampetti, Scalabrino, Oliveto, Canfora, and Di~Penta}]{zampetti2017open}
Zampetti F, Scalabrino S, Oliveto R, et~al (2017) How open source projects use static code analysis tools in continuous integration pipelines. In: Int. Conf. on Mining Software Repositories, pp 334--344

\bibitem[{{Zheng} et~al.(2006){Zheng}, {Williams}, {Nagappan}, {Snipes}, {Hudepohl}, and {Vouk}}]{Zheng2006}
{Zheng} J, {Williams} L, {Nagappan} N, et~al (2006) On the value of static analysis for fault detection in software. IEEE Transactions on Software Engineering 32(4):240--253

\end{thebibliography}

\end{document}